\documentclass[journal]{IEEEtran}
\ifCLASSINFOpdf
\else
\fi

\usepackage[switch]{lineno}

\usepackage[hidelinks]{hyperref}
\usepackage{graphicx}
\usepackage{subfig}
\usepackage{dblfloatfix}
\usepackage{array}
\usepackage{multirow}
\usepackage{algorithm}
\usepackage{algpseudocode}
\usepackage{booktabs}
\usepackage{tablefootnote}
\usepackage{xcolor}
\usepackage{amsmath}

\begin{document}
%
\title{Fed-LSAE: Thwarting Poisoning Attacks against Federated Cyber Threat Detection System via Autoencoder-based Latent Space Inspection}
%
%
%

\author{Tran Duc Luong\IEEEauthorrefmark{1}\IEEEauthorrefmark{2},
        Vuong Minh Tien\IEEEauthorrefmark{1}\IEEEauthorrefmark{2}, Nguyen Huu Quyen\IEEEauthorrefmark{1}\IEEEauthorrefmark{2}, \\Do Thi Thu Hien\IEEEauthorrefmark{1}\IEEEauthorrefmark{2}, 
          Phan The Duy \IEEEauthorrefmark{1}\IEEEauthorrefmark{2},
        Van-Hau Pham \IEEEauthorrefmark{1}\IEEEauthorrefmark{2}\\
\IEEEauthorblockA{\IEEEauthorrefmark{1}Information Security Laboratory, University of Information Technology, Ho Chi Minh city, Vietnam\\
\IEEEauthorrefmark{2}Vietnam National University, Ho Chi Minh city, Vietnam
\\
\{19521815, 19522346\}@gm.uit.edu.vn, \{quyennh, hiendtt, duypt, haupv\}@uit.edu.vn
}
\thanks{Tran Duc Luong, Vuong Minh Tien, Nguyen Huu Quyen, Do Thi Thu Hien, Phan The Duy, Van-Hau Pham are with Information Security Lab (InSecLab), University of Information Technology, Vietnam National University Ho Chi Minh City, Hochiminh City, Vietnam. Website: (see at http://uit.edu.vn).}
\thanks{The corresponding author is Van-Hau Pham (Email: haupv@uit.edu.vn).}
\thanks{Manuscript received April xx, 20xx; revised August xx, 20xx.}}

%
%

\markboth{Journal of \LaTeX\ Class Files,~Vol.~xx, No.~x, August~20xx}%
{Shell \MakeLowercase{\textit{et al.}}: Bare Demo of IEEEtran.cls for IEEE Journals}
%



\maketitle

\begin{abstract}
The significant rise of security concerns in conventional centralized learning has promoted federated learning (FL) adoption in building intelligent applications without privacy breaches. In cybersecurity, the sensitive data along with the contextual information and high-quality labeling in each enterprise organization play an essential role in constructing high-performance machine learning (ML) models for detecting cyber threats. Nonetheless, the risks coming from poisoning internal adversaries against FL systems have raised discussions about designing robust anti-poisoning frameworks. Whereas defensive mechanisms in the past were based on outlier detection, recent approaches tend to be more concerned with latent space representation. In this paper, we investigate a novel robust aggregation method for FL, namely Fed-LSAE, which takes advantage of latent space representation via the penultimate layer and Autoencoder to exclude malicious clients from the training process. The experimental results on the CIC-ToN-IoT and N-BaIoT datasets confirm the feasibility of our defensive mechanism against cutting-edge poisoning attacks for developing a robust FL-based threat detector in the context of IoT. More specifically, the FL evaluation witnesses an upward trend of approximately 98\% across all metrics when integrating with our Fed-LSAE defense.

\end{abstract}


%
\IEEEpeerreviewmaketitle

\section{Introduction} \label{sec: intro}
%
%
%
%
\
\IEEEPARstart{R}{ecently,} the Internet of Things (IoT) has emerged as a transformative technology that is reshaping the way we interact with the world around us. In fact, IoT refers to an expanding network of interrelated devices equipped with sensors, software, and additional technologies that facilitate data exchange among themselves and with other systems via the Internet. The potential applications of IoT are diverse and span a range of sectors including healthcare, transportation, agriculture, manufacturing, and smart cities. However, the rapid expansion of the IoT also poses significant security risks that must be addressed \cite{serror2020challenges_sur} \cite{Liu2019_secure_iot_Sur}. One of the most common threats that IoT systems have to face is cyberattacks. Cybercriminals can exploit security flaws in both IoT networks and devices in order to attain unauthorized access, exfiltrate sensitive information, cause physical damage, or carry out large-scale attacks such as distributed denial-of-service (DDoS). Therefore, the need for an intrusion detection system (IDS) as a layer of defense against cyber threats in IoT infrastructure is becoming more and more crucial \cite{Arisdakessian2023_iot_ids_sur}  \cite{heidari2022internet_ids_sur}. 

Moreover, in order to enhance the capability of detecting unknown malicious traffics and leverage the vast amounts of data generated by IoT devices or large networks, machine learning (ML) \cite{Ridwan2021_ML-network-sur} has been implemented in constructing robust IDS systems \cite{Sahani2023_ML-IDS-sur} \cite{Ayesha2021_ML-IDS-cybersec} \cite{ahmad2021_NIDS-ML-DL} \cite{duy2023investigating}. Traditionally, the common technique to build ML models is centralized, where all training data is collected and stored on a single server. However, this method is becoming impractical \cite{Khan2021_FL-IoT-Survey} \cite{drainakis2020_FLvsCL} in fact due to privacy and security concerns surrounding data collection. Many concerns have been raised about the confidentiality of data owners due to the potential of sensitive data being compromised or lost during the data storage, transmission, or sharing process. Additionally, the heavy computational cost is also a major challenge in training a conventional ML model. 

In this context, Federated Learning (FL) \cite{Fauzi2022_FLvsCL} has emerged as a new paradigm of distributed machine learning for building intelligent and privacy-preserving applications in IoT ecosystems and smart cities \cite{Nguyen2021_FL_iot_sur}, \cite{Venkatasubramanian_iotmal_fl_sur}. This learning scheme allows multiple devices or entities to train a shared model collaboratively while keeping the training data decentralized. To be more specific, after initializing the model, the global server transmits it to each participating collaborator. Then each device trains the model on its local dataset and sends the updated model parameters back to the central server for aggregation. This procedure is repeated in several rounds so that the model can gain knowledge from a variety of data sources while maintaining data privacy and confidentiality. Therefore, adopting FL \cite{Alazab2022_FL_4_cybersec_sur} \cite{Ghimire2022_FL_cybersec_sur} \cite{Mothukuri2021_FL-IoT-AnomalyDec} to train robust ML-based threat detectors for IoT would be a potential strategy when it comes to cybersecurity.

Nevertheless, FL systems have to deal with poisoning attacks \cite{Cao2022_ModelPoisonFL} \cite{TolpeginTGL20_poisonFL} \cite{Jiale2019_Poison-GAN-FL} \cite{Virat2022_poison_productionFL} from its internal parties since the central server has no right to access the private local data of its collaborators. Unfriendly participants might pretend to be honest clients and manage to corrupt the learning phase by injecting malicious data (data poisoning) \cite{TolpeginTGL20_poisonFL} or modifying updated model parameters (model poisoning) \cite{fang2019_LMPA} \cite{Cao2022_ModelPoisonFL}. In this way, they could deteriorate the general performance of the global model (untargeted \cite{Mallah2021_untargetedPA} \cite{Virat2022_poison_productionFL}) or make a bias in predicting attacker-chosen class inputs (targeted \cite{Jebreel2022_FLDefender-Targeted} \cite{andreina2020_baffle}). The targeted poisoning attacks are more advanced and trickier to carry out since it requires the stealthiness of adversaries to be undetected by threat hunters while maintaining the original detecting function on the remaining classes.

As a consequence, numerous robust aggregation schemes \cite{Jebreel2022_FLDefender-Targeted} \cite{andreina2020_baffle} \cite{Xingyu2023_LoMar} \cite{Jiang2023_FL_Def_LF} \cite{Zhang2022_SecFedNIDS} against poisoning attacks have been introduced to the research community in the past few years. Almost of defense mechanisms \cite{NCVy2021_FL-Poison-Defense} \cite{Zhang2022_SecFedNIDS} \cite{kim2022_backdoor-def-FL} \cite{lai2023two_poison_FL_ids} are based on anomaly detection and verify an anomalous updated model as an outlier compared to benign groups. Outliers then are excluded from the server-side aggregation to ensure the stable performance of the global model. However, these methods rely on the model parameter space, while ML-based architectures in reality are constructed from thousands to millions of parameters. Therefore, the mentioned defensive frameworks must deal with a plenty of huge-sized weight parameters, posing a heavy burden on computational cost as well as the difficulty in clearly detecting poisonous models. In addition, some defense methods \cite{Zhang2022_SecFedNIDS} also need to determine the number of attackers in advance, causing impracticality in real-world environments. At the same time, these solutions encounter numerous obstacles in differentiating between malicious model parameters and benign ones trained on non-independent and identically distributed (non-IID) data \cite{zhu2021_FL_NonIID_Sur} \cite{li2021_FL_NonIID_Silos} \cite{Xuming2022_FL_NonIID_Sur}.

A new approach using latent space like FedCC \cite{Jeong2022_FedCC}, FLARE \cite{wang2022_flare} has been published recently to address those shortcomings. By extracting and comparing Penultimate Layer Representations (PLRs) among models, the solutions of FedCC \cite{Jeong2022_FedCC} and FLARE \cite{wang2022_flare} have shown that PLRs of benign model parameters follow an identical pattern, while poisonous PLRs adhere to other directions. However, \cite{wang2022_flare} requires an auxiliary dataset in the server to compute PLR for each local model, which violates the FL standards of data privacy, especially in the case of a compromised global server. Meanwhile, the FedCC approach\cite{Jeong2022_FedCC} extracts PLRs directly from the updated models without using an auxiliary dataset, which could cause the instability of PLRs if each local model was trained on a different data distribution.

Thus, in this paper, we propose a new latent space (LS)-based anti-poisoning mechanism called Fed-LSAE to tackle poisoning attacks in FL-based systems, even in non-IID data environments. More specifically, Penultimate Layer Representation (PLR) would be utilized as the first LS-based core component in detecting malicious models. Moreover, to address the mentioned issue in FedCC \cite{Jeong2022_FedCC}, we implement Autoencoder (AE) as a second module to reduce the uncertainty of PLR changing and extract the LS representation of PLRs through the bottleneck layer assuming that PLR's parameter size is still massive. By learning the latent space of updated models, we can recognize the similarity level between updated weight models and the global model via Centered Kernel Alignment (CKA) algorithm. The CKA scores are then clustered into two groups, where the larger cluster of models will be selected as benign ones and utilized for the FedAvg aggregation.

To sum up, we outline the primary contributions of this study as follows.
\begin{itemize}
    \item This work deeply investigates three typical types of poisoning attacks, including both data and model poisoning, against FL-based threat detectors in the context of IoT. From that, we designed a new anti-poisoning scheme named Fed-LSAE by detecting malicious uploaded weight parameters via Autoencoder-based latent space representations. Especially, this approach do not need any prior knowledge or raw data on the server like previous works \cite{Zhang2022_SecFedNIDS} \cite{wang2022_flare} \cite{NCVy2021_FL-Poison-Defense}. 
    \item We conduct several experimental scenarios to reveal the effectiveness of our defense against poisoning attacks through an in-depth analysis of two datasets of IoT cyberattacks with different ML models. The proposed approach can work well when the percentage of adversaries is up to 40\%. 
    \item By integrating AE into the defense mechanism, we show that our Fed-LSAE can outperform FedCC \cite{Jeong2022_FedCC} in clearly distinguishing between benign clients (having IID or non-IID data) and malicious ones. In addition, our proposed framework is more conducive to identifying and removing poisoned updates from the model aggregation stage, even in non-IID settings.
\end{itemize}

The remaining sections of this article are constructed as follows. Section \ref{related_work} introduces some related works in poisoning attacks against FL-based models and the countermeasures. The following Section \ref{Background} gives a brief background of applied components. Next, the threat model and methodology are discussed in Section \ref{methodology}. Section \ref{experiments} describes the experimental settings and scenarios with result analysis of our Fed-LSAE performance. Finally, we conclude the paper in Section \ref{conclusion}.


\section{Related work} \label{related_work}

\subsection{Poisoning Attacks in the context of FL}
Despite providing a privacy-preserving training mechanism, FL still exposes many vulnerabilities that can be exploited by adversaries in multiple ways \cite{rodriguez2023survey_fl_threats}. Poisoning attacks are one of the most common techniques that can be easily conducted to devastate FL training performance. Based on the attacker's strategies, poisoning attacks can be separated into data poisoning and model poisoning. The former occurs when attackers try to manipulate their own data with the aim of updating malicious model parameters, resulting in disrupting the FL model. More specifically, adversaries could conduct label-flipping techniques or inject perturbed samples into the local training data to achieve their goals. Meanwhile, model poisoning is a type of attack where unfriendly clients directly modify the weight of updating models during the training process that changes the decision boundary of the model, causing it to classify certain inputs differently than it would have otherwise or even hindering convergence. In addition, adversarial clients can manipulate certain settings of the model during the training process, such as learning rate, the number of epochs for local training or batch size, etc. In general, model poisoning attacks are regularly easier to conduct and more efficient than data poisoning ones because they do not focus on data preparation but manipulate the weight parameters which might directly influence the productivity of the global aggregation. 

Numerous published recent works \cite{Jiale2021_poisonGAN} \cite{Jiale2019_Poison-GAN-FL} \cite{Wang2020_FL-Backdoor} \cite{andreina2020_baffle} have proved the efficiency of poisoning attacks in exerting a significant impact on the FL performance. To be more specific, Jiale Zhang et al. \cite{Jiale2019_Poison-GAN-FL} proposed a GAN-based poisoning attack strategy against the federated image classifier. Attackers pretend to be reliable participants so that they can utilize the global model as a GAN discriminator to mimic other participants’ training samples from a noise vector. Through evaluation of MNIST and AT\&T datasets, this paper has shown that FL would be vulnerable to adversarial poisoning attacks in which any internal party has updated local model parameters trained on poisoned data to the aggregation server. Developed from the above article, Jiale Zhang et al. \cite{Jiale2021_poisonGAN} also presented a generative poisoning mechanism named PoisonGAN against the FL protocol in the context of edge computing. This paper built two types of poisoning attack techniques: backdoor and label-flipping, to assess the feasibility of the adversarial poisoning attack against the FL framework in practice. Furthermore, Sebastien Andreina et al. \cite{andreina2020_baffle} also examined the efficiency of model poisoning attacks against FL-based system by conducting a backdoor attack based on the principle of multitask learning: the backdoor samples train the local model on the adversarial subtask while the genuine ones help preserve behavior of model on the primary task. The authors indicated that this attack strategy can be destructive even when existing only one poisoned update in a single round. A defense framework, named BaFFLe was also published to detect backdoor attacks on CIFAR-10 and FEMNIST dataset in their work. 

\subsection{Defense mechanisms against poisoning attacks in FL}\label{subsec:defense}

To mitigate the risk of poisoning attacks in FL, researchers have proposed various defense mechanisms \cite{Jiang2023_FL_Def_LF} \cite{Jebreel2022_FL_Def_LF} \cite{Awan2021_CONTRA} \cite{Xingyu2023_LoMar} \cite{Zhang2022_FLDetector}. These techniques aim to detect and reduce the effects of poisoned local updates on the global model by identifying the malicious clients and removing their updates from the learning process.

A familiar defense technique in previous anti-poisoning works is adopting outlier detection algorithms to reveal poisoned updates as outliers and remove them from aggregation. For instance, Nguyen Chi Vy et al. \cite{NCVy2021_FL-Poison-Defense} investigated the federated IDS performance when conducting label-flipping and adversarial attacks on the Kitsune dataset. A new anti-poisoning scheme was introduced, which uses the Local Outlier Factor (LOF) algorithm to verify local updated parameters from internal collaborators. By computing the LOF distance score between the uploaded model weights and the benign history, it can reveal whether an updated local model belongs to a malicious agent or not. Although this framework showed a great defensive performance against poisoning attacks, it must ensure that the FL system starts with only benign updates in several rounds. Also, the problem of non-IID data was not discussed in the paper. 

In recent times, Yuan-cheng Lai et al. presented DPA-FL \cite{lai2023two_poison_FL_ids} framework as a two-phase defensive mechanism against label flipping and backdoor attacks in the context of FL-based IDS. Specifically, DPA-FL also adopted the LOF algorithm as the first stage, called relative phase, to discriminate obvious malicious models from benign ones through the significant difference of LOF anomaly scores. Towards some local models with middle LOF scores, they have to undergo the second phase (absolute phase) for further data testing to confirm. However, the experimental results on the CICIDS2017 dataset only showed its effectiveness in the case of IID data. We can see that if there exists collaborative agents with non-IID data, the relative phase using LOF will be ineffectual when it comes to clarifying the difference between benign non-IID weight parameters and malicious ones. In other words, benign non-IID models might obtain the same LOF anomaly score as poisoned ones. 

Meanwhile, the study \cite{Zhang2022_SecFedNIDS} also proposed the robust defense against label flipping and clean label attacks for FL-based network IDS, namely SecFedNIDS, which consists of 2 defense stages: model-level and data-level. Firstly, the Stochastic Outlier Selection (SOS) algorithm is applied to the model-level defensive mechanism at the server side. Its goal is to detect poisoned updated models as outliers based on the relationship among the uploaded local model parameters, and then reject them from the global aggregation. Nevertheless, this SOS-based method needs to know the number of attackers in advance, which seems infeasible in the real-world context. Secondly, at the data-level stage, they propose a novel poisoned data detection approach based on class path similarity, in which the class path is retrieved by the layer-wise relevance propagation (LRP) algorithm. However, this method works only if there exists any interventions in local agents' datasets, which can lead to another threat called inference attack \cite{Lingjuan2020_ThreatFL} \cite{nasr2019_inference-attack} \cite{fu2022_inferencelabel} or even break the rules of privacy preservation in FL. Additionally, these papers \cite{NCVy2021_FL-Poison-Defense} \cite{lai2023two_poison_FL_ids}\cite{Zhang2022_SecFedNIDS} only work in the model parameter space, which puts a burden on computational costs and resource consumption. 

Recently, there has been renewed interest \cite{Jeong2022_FedCC} \cite{wang2022_flare} in using latent space representation to build defensive schemes for FL-based systems against model poisoning attacks (MPAs). Ning Wang et al. were the pioneer of this trend when discovering a robust model aggregation mechanism for FL, namely FLARE \cite{wang2022_flare}. FLARE leveraged penultimate layer representation (PLR) of models to differentiate malicious models from benign patterns. By extracting PLR of each model through an auxiliary data, FLARE determined a trust score for each local model based on pairwise PLR discrepancies among all updated models. As a result, the server aggregation could alleviate the impact of poisonous updates with low trust scores. Although FLARE could outperform some previous defenses, for example FLTrust \cite{cao2020fltrust}, in both IID and non-IID data cases, it still exposes some limitations such as the risk of data leakage when using an auxiliary dataset in the server, only accuracy metric was used to evaluate the performance of FLARE against untargeted MPAs. 

Furthermore, the likewise approach was proposed by Hyejun Jeong et al. with a defensive mechanism called FedCC \cite{Jeong2022_FedCC}. While FedCC does not require any subset of raw data or information sharing to extract PLRs, it compared the similarity between each PLR of local update and PLR of the global model by CKA algorithm. The lower the CKA score is, the higher the likelihood that it is a poisoned model. The experimental results on three datasets indicated that FedCC surpassed FLARE \cite{wang2022_flare} in all scenarios in detecting the state-of-art MPAs. However, retrieving PLR from updated models directly without the same dataset can lead to the uncertainty of PLR vectors, which might pose a negative impact on the poisoning detection rate. 

Lately, Yifeng Jiang et al. \cite{Jiang2023_FL_Def_LF} presented MCDFL, a detection mechanism against label flipping attacks via data quality inspection. The server-side pretrained generator is delivered to each local agent to extract latent feature space as data quality according to the given label sequences. By updating these data quality metrics, the server can clarify malicious data distribution from benign patterns via K-means clustering algorithm. The success rate of MCDFL method, however, is not always guaranteed since the data quality extraction process is conducted on the client-side. As a result, the adversaries can adapt to make some perturbations on its updated data quality parameters. Also, the feasibility of MCDFL is not discussed in other advanced data poisoning attacks.
 
To reduce the aforementioned limitations in previous works, we propose a Fed-LSAE module as a latent space-based defensive framework against different types of poisoning attacks, even in non-IID settings. Our Fed-LSAE does not require any prior knowledge or datasets for the poisoning detection on the server-side. Not only can Fed-LSAE protect FL systems from model poisoning attacks, but our recommended approach is also effective in detecting data poisoning attacks by absorbing data representation via PLR vectors. In addition, the Fed-LSAE could address the PLR instability issue by implementing AE to learn the benign pattern of PLR vectors before the training process.

\section{Background} \label{Background}
\subsection{Penultimate Layer Representation}
\begin{figure}[!t]
    \centering
    \includegraphics[width=0.25\linewidth]{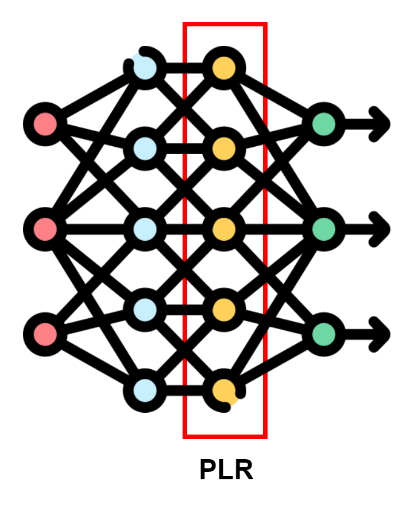}
    \caption{The penultimate layer in a neural network.}
    \label{fig:PLR}
\end{figure}

The penultimate layer refers to the second-to-last layer of a neural network that is just before the output layer (\textbf{Fig. \ref{fig:PLR}}). The Penultimate Layer Representation (PLR) is a vector of numbers that encodes the input data into a feature space optimized for the specific task the neural network attempting to perform. The output layer of the network then uses this feature vector to make its final predictions or classifications. In other words, we can learn the input data representation via PLR.

In \cite{wang2022_flare} and \cite{Jeong2022_FedCC}, authors demonstrated that benign PLRs follow the same distribution while malicious ones stick to other directions. Additionally, \cite{Jeong2022_FedCC} indicated the penultimate layer is the most distinct layer out of all layers in the neural networks, which means we can classify local models in FL via PLR instead of all model parameters.

\subsection{Centered Kernel Alignment}

The Centered Kernel Alignment (CKA) algorithm was introduced by Kornblith et al. \cite{Kornblith2019_CKA} to compare feature representations in neural networks. It is designed to measure the similarity between the representations by aligning their respective kernel matrices. In a normalized version, the CKA score is computed based on the Hilbert-Schmidt Independence Criterion (HSIC) as in \textbf{Eq.~(\ref{eq: cka})}. 

\begin{equation} \label{eq: cka}
CKA(K, L) = \frac{HSIC(K, L)}{\sqrt{HSIC(K, K)HSIC(L, L)}}
\end{equation}

where $K$ and $L$ are kernel matrices corresponding to two feature representations. The resulting CKA score ranges from 0 and 1, in which a score of 1 indicates perfect alignment between the two sets of feature representations. 

In this work, we utilize CKA as a benchmark to evaluate the resemblance between each local LS representation and the global one. We can filter malicious LS representations which are likely to be distinct from the global LS. Compared to other algorithms such as cosine, CKA could offer a more evident difference between a malicious model and a non-IID-based model when comparing the similarity with a benign one.

\subsection{Latent space representation in Autoencoder}

\begin{figure}[b]
    \centering
    \includegraphics[width=0.8\linewidth]{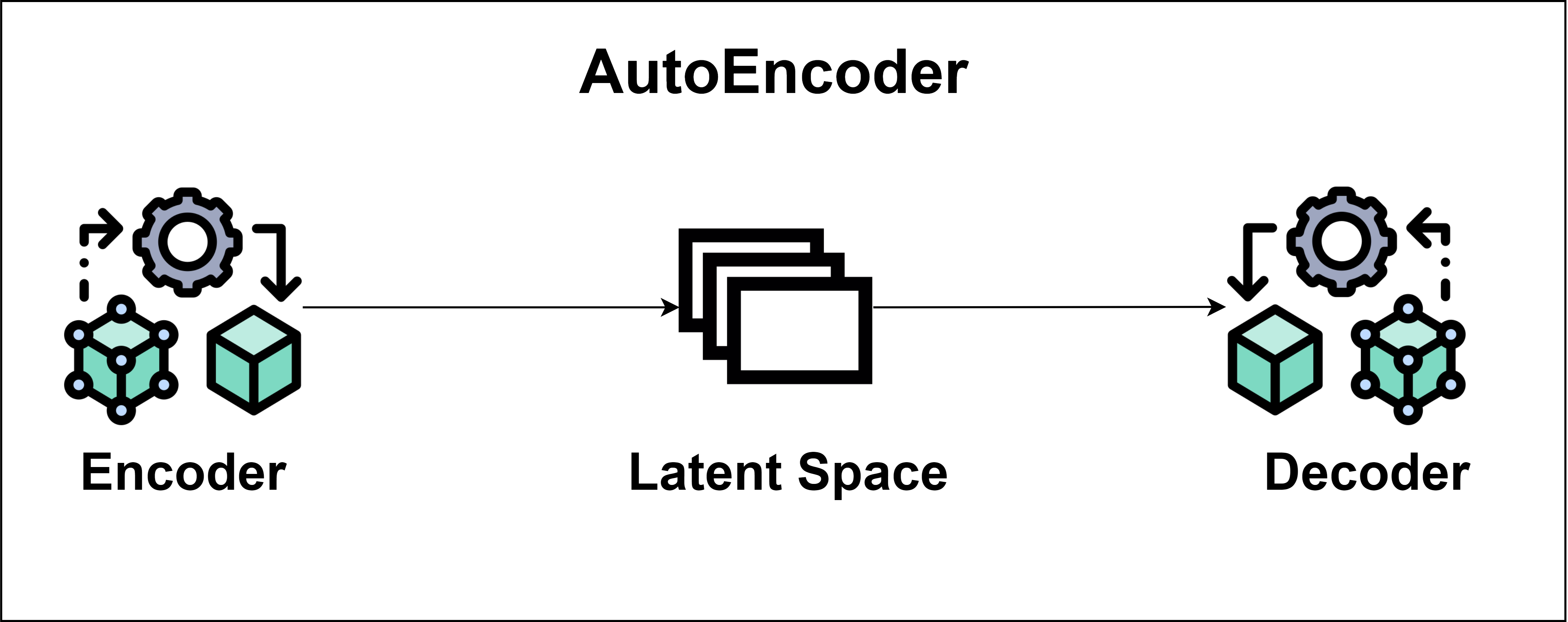}
    \caption{The Autoencoder (AE) architecture.}
    \label{fig:ae_arc}
\end{figure}

Autoencoder (AE) is an unsupervised ML algorithm which is leveraged to represent data for the task learning. As shown in \textbf{Fig.~\ref{fig:ae_arc}}, it consists of two main components: an Encoder and a Decoder. The encoder compresses input data into a lower-dimensional representation, while the decoder reconstructs that representation to the original data shape.

The above-mentioned compressed representation is latent space representation in AE. By learning the latent space representation of input data, AE can capture the most important features or patterns in the input. As a result, it can be applied in various fields, such as dimensionality reduction, feature extraction, anomaly detection, etc. Moreover, another benefit of AE is the capability of learning to represent complex data distributions with a relatively small number of parameters. This means that even a small portion of data can be sufficient to train an AE effectively, as long as the data is representative of the underlying data distribution. In this paper, by feeding PLR vectors into a pre-trained AE, we can retrieve their latent space representations, which are then used in detecting malicious models.


\section{Methodology} \label{methodology}

 \subsection{Threat Model}

\subsubsection{Threat Model}\hspace*{\fill} 

For this study, we assume that the number of adversaries is always less than half of the total number of clients, 40\% more precisely. The remaining participants and the server are considered trusted parties during the process of training global model. Meanwhile, the attacker nodes continuously carry out poisoning attacks against the FL system, which means such nodes constantly update their poisoned local model.

\subsubsection{Attacker's Knowledge and Ability}\hspace*{\fill} 

In the context of poisoning attacks, adversaries pretend to be benign participants with malevolent objectives in the FL framework. They have an in-depth insight into the training architecture since all the parties in collaborative learning adhere to a common learning algorithm, dataset, and model hyperparameters in advance. In other words, poisoning attacks in this work would be conducted in a white-box manner. In this section, we also define the capabilities of attackers in corrupting the global model as follows.

\textbf{Permitted.} The poisoners take absolute control of the local training procedure with their dataset. They can arbitrarily change some hyperparameters of the retrieved model from the global server so that poisoning attacks can achieve high performance.

\textbf{Not permitted.} Malicious participants have no right to interfere with the learning phase or training data of other participants. Moreover, they could not influence the server-side aggregation or modify the previously agreed-upon training algorithm.

\subsubsection{Attack Strategy}\hspace*{\fill}\label{subsec: attack_strategy}

\textbf{Data manipulation using Label Flipping.} This is a type of poisoning attack that aims to undermine the performance of the federated model by intentionally flipping the labels of some data samples used for training. In this work, we only perform a binary classification task, where the detector recognizes the 1-labeled examples as attacks and the 0-labeled ones as benign. Therefore, the number of adversaries would flip all labels to the opposite ones so that their local parameters would become poisonous to the convergence of the global model.

\begin{figure}[!t]
    \centering
    \includegraphics[width=1\linewidth]{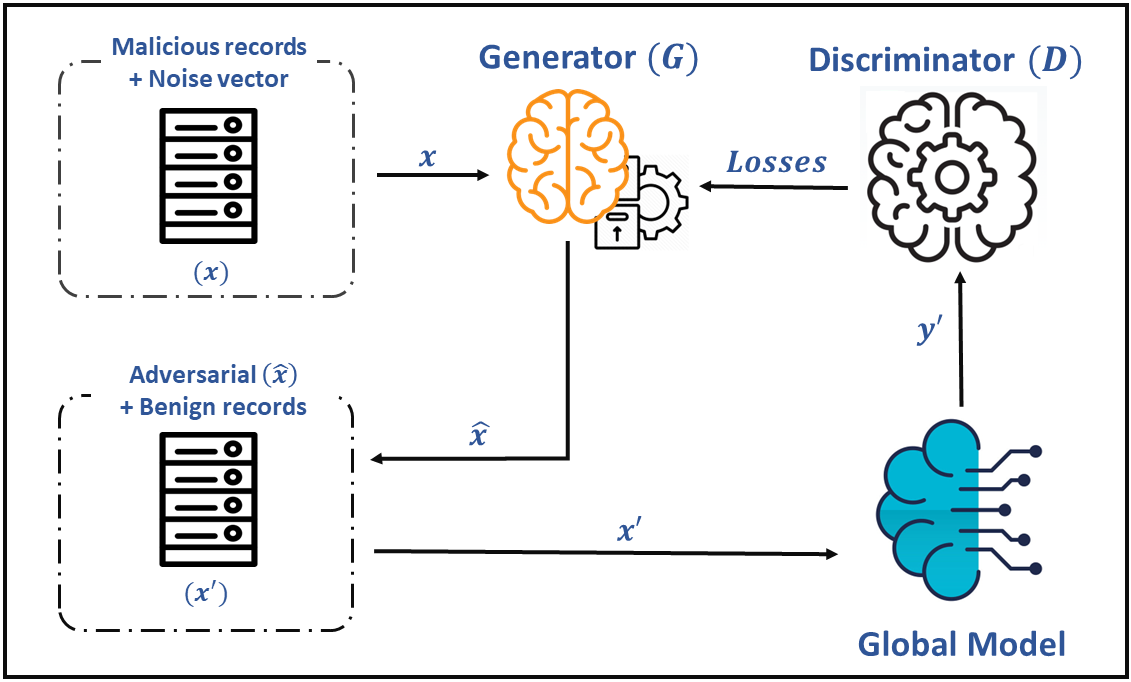}
    \caption{The GAN-based architecture for generating adversarial samples for poisoning attacks.}
    \label{fig:idsgan}
\end{figure}

\textbf{Data manipulation using GAN-based adversarial samples.} We leverage the IDSGAN \cite{Lin2022_IDSGAN} as the main GAN architecture in crafting adversarial samples to conduct data poisoning attacks. In other words, each unfriendly participant could train their own IDSGAN, as described in \textbf{Fig. \ref{fig:idsgan}}. The global model plays a role as the IDS component in IDSGAN. By feeding malicious samples into IDSGAN, attackers generate adversarial ones to train their poisoned local model, which is conducive to the misclassification of the global model. 

\textbf{Weight-scaling Model Poisoning.} Model poisoning attacks can be classified into two categories based on the attacker's objectives: untargeted and targeted. In untargeted attacks, the attacker's goal is to reduce the overall accuracy of the model, whereas in targeted attacks, the objective is to manipulate the model into misclassifying a particular class of inputs. The former aims to make the model less effective in general, while the latter seeks to create a specific bias in the model's decision-making process. In this paper, we only focus on the untargeted approach, where adversaries try to scale up their model weights $\alpha$ times before transmitting it to the aggregation server. \textbf{Eq.~(\ref{eq:mp})} has shown the model weights $w_{i}$ of the $i$-th client as an attacker after scaling up its original model. 

\begin{equation}\label{eq:mp}
        w_{i} \gets \{\alpha w_i^1, \alpha w_i^2, ..., \alpha w_i^P\}
\end{equation}

where $w_i^p$ refers to the $p$-th parameter value of $w_{i}$, and $P$ is the total number of model parameters.

\subsection{Detailed design of Fed-LSAE}
\begin{figure*}[!t]
    \centering
    \includegraphics[width=0.75\linewidth]{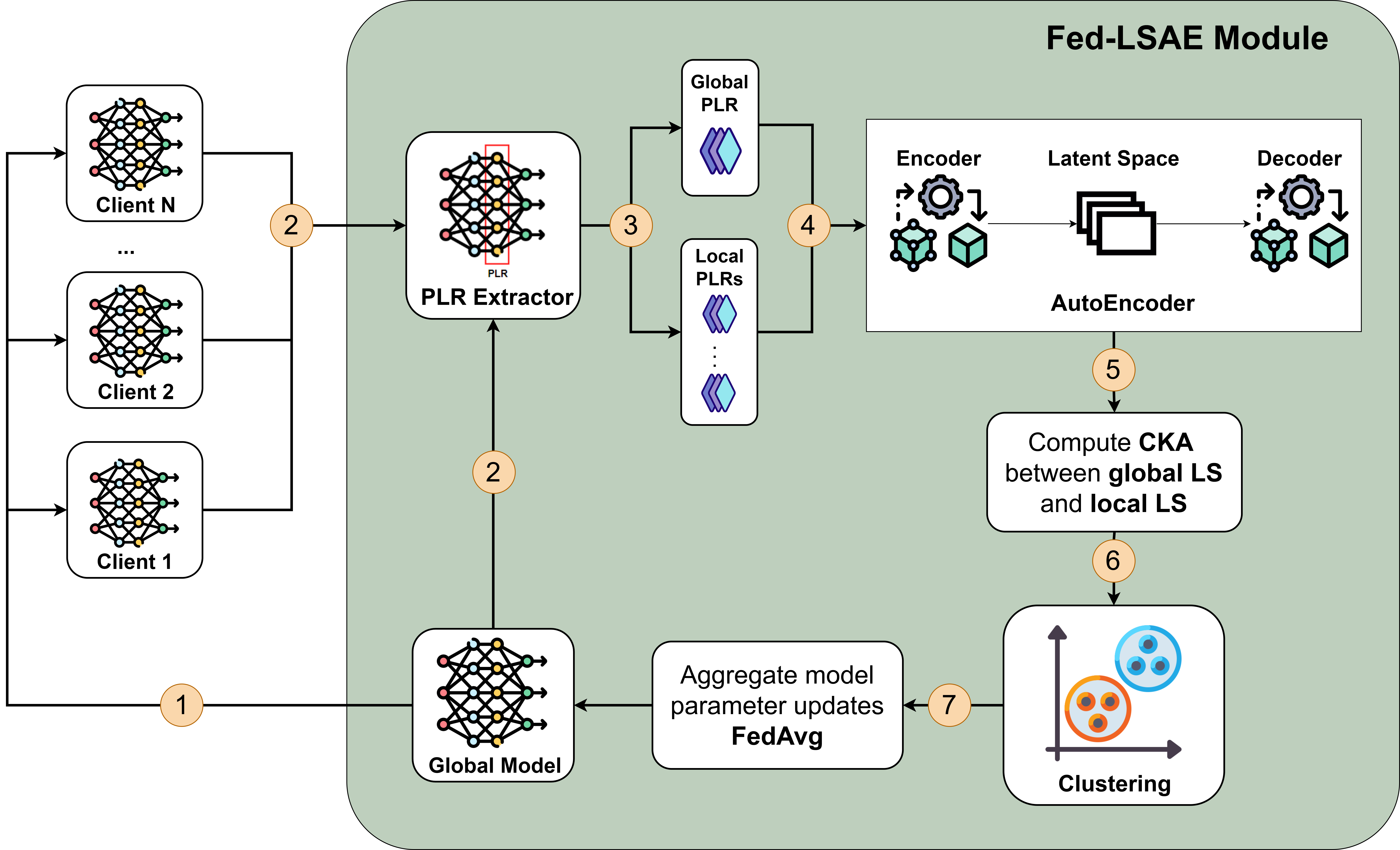}
    \caption{The Fed-LSAE architecture for Federated Threat Detection System based on Latent Space Representations.}
    \label{fig:Fed-LSAE}
\end{figure*}
The overall architecture of our proposed system is given in \textbf{Fig.~\ref{fig:Fed-LSAE}}.
\subsubsection{Architectural components}
\paragraph{Training clients} They are local agents which train ML-based models with their dataset before sending model weights to the aggregation server for computing the global model.

\paragraph{Aggregation server} The Fed-LSAE is located on the aggregation server to detect and remove malicious updates from the global training. It consists of 3 elements, including:

\begin{itemize}
    \item PLR Extractor: This module is responsible for extracting the PLR sequence of the inputs, which are the updated local models and global model. 
    \item Pretrained Autoencoder (AE): Via feeding PLR vectors into AE, it produces compressed representations containing the most important features of each PLR. The more detailed information of this element is depicted in \textbf{Section \ref{subsec: ae-pretrain}}.
    \item Clustering algorithm: By clustering CKA scores into two groups, we can indicate the smaller group as adversarial members and then filter them from the federated training process.
\end{itemize}

\subsubsection{Pretraining process of Autoencoder} \label{subsec: ae-pretrain} 
Before transmitting a duplicate version of the global model to participating agents, the global model is collaboratively trained in a few internal server-side organizations by their dataset in one round. We assume that all of these organizations are friendly, benign and belong to the global server so that AE can learn the characteristics of benign models in over $e$ epochs. Thereby, the encoder takes a PLR vector $x$ of each benign aforementioned model as input, and produces a latent space representation $h$ using a series of nonlinear transformations as follows:

\begin{equation} \label{eq:encoder}
h = f_{\theta_1}(x)
\end{equation}
where $f_{\theta_1}$ represents the encoder function with learnable parameters $\theta_1$. On the other hand, the decoder takes the latent representation $h$ and produces a reconstructed output vector $\hat{x}$ (\textbf{Eq.~(\ref{eq:decoder})}).

\begin{equation} \label{eq:decoder}
\hat{x} = g_{\theta_2}(h)
\end{equation}
where $g_{\theta_2}$ represents the decoder function with learnable parameters $\theta_2$. The objective of the autoencoder is to minimize the difference between the input vector $x$ and the reconstructed output vector $\hat{x}$. Therefore, we make use of Mean Squared Error (MSE) in \textbf{Eq.~(\ref{eq:mse})} as the reconstruction loss.

\begin{equation}\label{eq:mse}
    L(x,\hat{x})=\frac{1}{n} \sum\limits_{i=1}^n (x_i-\hat{x}_i)^2
\end{equation}
where $x_i$ and $\hat{x}_i$ represent the $i$-th element of the input and reconstructed data, respectively.

We choose AE because it is an unsupervised algorithm that can learn inputs' representation effectively even in the case of limited training data points. Also, another benefit of AE is its capability of learning non-IID patterns if the internal server-side organizations have various data distributions. As a result, it fosters the accuracy of the following CKA calculation phase that reduces the likelihood of misclassifying benign non-IID models as malicious ones. 

\subsubsection{Workflow of Fed-LSAE}
\textbf{Fig.~\ref{fig:Fed-LSAE}} also illustrates the workflow of Fed-LSAE module integrated into the server-side process of FL-based threat detectors. To be more specific, the training procedure undergoes the following steps.

\begin{itemize}
    \item \textit{Step 1}: Initially, the central server initializes a new global model and an Autoencoder (AE) architecture, which then simultaneously experiences an internal training process on the server side (described in \textbf{Section \ref{subsec: ae-pretrain}}) to learn important features of benign PLR vectors. The resulting global model is then delivered to selected $k$ out of $n$ collaborative agents.
    \item \textit{Step 2}: The total $k$ clients train each local model on their own dataset, and then update the trained model weights to the server for aggregation. It is time for the Fed-LSAE module to make the action. The updated weight parameters from $k$ agents, as well as the global weight, are sent directly to a PLR extractor.

\begin{algorithm}[t]
\caption{The mechanism of Fed-LSAE for thwarting poisoning attacks.}\label{algo:fed-lsae}
\begin{algorithmic}[1]
  \Require  Global weight $W$,
        \Statex \hspace*{1.2em} A set of $n$ local weights $L=\{w_1,w_2,...,w_n\}$,
        \Statex \hspace*{1.2em} A pretrained AE model $AE$.
  \Ensure Aggregated global weight $W'$ 
\Procedure{FED-LSAE}{$W$, $L$}
  \State\Comment{\textbf{Extract PLRs of $W$ and $L$ \phantom{aaaaaaaaaaaaaaaaaa}}}
  \For{$i < n$}  
    \State $local\_plr[i]\gets L[i][penultimate\_layer]$;
  \EndFor 
  \State $global\_plr\gets W[penultimate\_layer]$;
  \State\Comment{\textbf{Compute AE-based latent space (LS) of PLRs \phantom{a}}}
  \For{$i < n$}  
    \State $local\_ls[i]\gets  AE.Encoder(local\_plr[i])$;
  \EndFor
  \State $global\_ls\gets AE.Encoder(global\_plr)$ 
  \State\Comment{\textbf{Calculate CKA score between global LS and each local LS \phantom{aaaaaaaaaaaaaaaaaaaaaaaaaaaaaaaaaaaaaaa}}}
  \For{$i < n$}  
    \State $cka[i]\gets  CKA(global\_ls, local\_ls[i])$;
  \EndFor
  \State\Comment{\textbf{Cluster CKA scores into 2 groups\phantom{aaaaaaaaaaaaa}}}
  \State $result = cluster\_model (n\_clusters=2, cka)$;
  \State $poisoned\_group\gets result.smaller\_group$;
  \State $benign\_group\gets result.larger\_group$;
  \State\Comment{\textbf{Aggregate new robust global weight \phantom{aaaaaaaaaa}}}
  \State $W'= FedAvg(benign\_group)$;
  \State \textbf{return} $W'$;
\EndProcedure
\end{algorithmic}
\end{algorithm}
    \item \textit{Step 3}: As described in \textbf{Algorithm \ref{algo:fed-lsae}}, the PLR extractor outputs the global PLR for the global model and $k$ local PLRs (lines 3-6) to feed into the pre-trained AE model.

    \item \textit{Step 4}: Later, the latent space representation (LSR) of each PLR will be retrieved via the bottleneck layer of the AE (lines 8-11). As mentioned before, this step aims to reduce the instability of PLR vectors if local agents train their models in different data distributions. Furthermore, it can minimize the computational costs on the assumption that the PLR dimension is still relatively large. 
    \item \textit{Step 5}: In this step (lines 13-15), we leverage the Radial Basis Function (RBF) CKA algorithm to measure the similarity level between the global LSR and each local LSR. The reason for using RBF-CKA is that it can show the similarity differences among non-IID models are slighter than the similarity differences between non-IID models and malicious ones. Note that, non-IID models here are trained by benign clients, in which each of them has a different data distribution in terms of the number of data points or the ratio between normal and attack samples.
    \item \textit{Step 6}: This stage involves gathering CKA scores into two groups by using a clustering algorithm (lines 17-19). In our work, K-means is used for this task. Since the number of attackers cannot exceed 50\% of total clients, we assign the greater cluster as benign members, while the other will be poisonous models.
    \item \textit{Step 7}: This is the final step where the benign group is selected for the FedAvg aggregation (line 21). In other words, adversaries with poisoned updates would be removed from aggregating a new version of the global model, which results in the robust aggregation for the FL system. The FedAvg algorithm is defined as \textbf{Eq. (\ref{eq:fedavg})}.
    
    \begin{equation}\label{eq:fedavg}
        w_{t+1} \gets \sum_{i=1}^k \frac{n_i}{n} w_{i, t}
    \end{equation}

    where $w_{t+1}$ is the updated global model at round $t+1$, $w_{i, t}$ is the local model of the $i$-th benign client at round $t$, $n_i$ is the number of local data points of the $i$-th benign client, $n$ is the total number of local data points across all clients in the selected benign cluster, and $k$ is the total number of benign clients participating in the federated learning process.
    
\end{itemize}

The resulting version of the global model is then sent to newly selected $k$ agents, and this process is repeated from step 2 to step 7 until the global model obtains the convergence point.

\section{Experiments and Analysis} \label{experiments}

\subsection{Dataset and Preprocessing}
To conduct experiments on IoT network traffic attacks, we utilize recent ML-based NIDS datasets called CIC-ToN-IoT and N-BaIoT. 
\subsubsection{CIC-ToN-IoT}

CIC-ToN-IoT is a network traffic collection, extracted from the PCAP files of the ToN-IoT dataset by the CICFlowMeter-v4 tool. It contains more than 5.3 million network records with 85 features in a csv file, including roughly 53\% attack instances and 47\% benign ones. The attack samples can be further classified into 9 cyberattack types including Backdoor, DoS, DDoS, Injection, etc. In our study, we only select 1,070,158 samples to conduct experiments while maintaining the proportion between benign examples and attack ones as mentioned.

Initially, there are 85 features for each record with 83 main features, a Label column (defining benign samples as 0 and attack samples as 1), and an Attack column (defining the types of attack). Due to our scope of binary classification, the Label column is used as the training target. In addition, we remove 14 redundant features with unique values or serve no purpose in labeling samples, such as Flow ID, Src IP, Dst IP, etc. The resulting training data has 70 dimensions along with the Label column. Moreover, any records containing non-numeric values (NaN) or infinity values (Inf) are also discarded. Finally, we apply a Min-max normalization as in \textbf{Eq.~(\ref{eq:minmax})} to the remaining 70 features to have their values in the range of [0,1].

\begin{equation}\label{eq:minmax}
    x_{scaled} = \frac{x - \text{min}(x)}{\text{max}(x) - \text{min}(x)}
\end{equation}

where $x_{scaled}$ is the normalized version of feature value $x$. $\text{max}(x)$ and $\text{min}(x)$ refer to the maximum and the minimum values of this feature in the dataset, respectively.

\subsubsection{N-BaIoT} 

The N-BaIoT dataset \cite{nbaiot} is a publicly available dataset released in 2019, designed for research on intrusion detection systems for IoT devices. It contains network traffic data from a heterogeneous IoT environment with 50 different types of devices, with both benign and malicious traffic data, and various types of attacks. In our experiments, we also take a subset of N-BaIoT to evaluate our Fed-LSAE which consists of over 800,000 samples with the ratio of benign instances and malicious ones is approximately 1:10.  

In these experiments, N-BaIoT undergoes the same preprocessing steps as CIC-ToN-IoT. The resulting dataset contains records with 115 features and a Label column with binary values of 0 and 1 for benign and attack samples, correspondingly. In addition, all 115 features are normalized to the range of [0,1] using the same Min-Max normalization as in \textbf{Eq.~(\ref{eq:minmax})}.

After the preprocessing phase, both datasets are divided into 3 parts for different usage.

\begin{itemize}
    \item Part 1 (70\%): the training dataset divided for $n$ participating agents.
    \item Part 2 (25\%): the testing dataset at the server-side to evaluate the global model performance.
    \item Part 3 (5\%): the training dataset that is evenly divided for internal server-side organizations with the aim of training AE in the initialization stage.
\end{itemize}

\subsection{Performance Metrics}

We evaluate our proposed method via 4 following metrics: Accuracy, Precision, Recall, F1-Score. Since our work conducts experiments in binary classification tasks, the value of each metric is computed based on a 2D confusion matrix which includes True Positive (TP), True Negative (TN), False Positive (FP) and False Negative (FN).

\textit{Accuracy} is the ratio of correct predictions $TP,~TN$ over all predictions. Mathematically, the \textit{Accuracy} of a model is calculated as \textbf{Eq.~(\ref{eq:acc})}.

\begin{equation}\label{eq:acc}
Accuracy = \frac{TP + TN}{TP + TN + FP + FN}
\end{equation}

\textit{Precision}, as in \textbf{Eq.~(\ref{eq:pre})},  measures the proportion of $TP$ over all samples classified as positive. 

\begin{equation}\label{eq:pre}
Precision = \frac{TP}{TP + FP}
\end{equation}

\textit{Recall}, which is defined in \textbf{Eq.~(\ref{eq:recall})}, measures the proportion of $TP$ over all positive instances in testing dataset.

\begin{equation}\label{eq:recall}
Recall = \frac{TP}{TP + FN}
\end{equation}

\textit{F1-Score} is the Harmonic Mean of $Precision$ and $Recall$ that has the formula as in \textbf{Eq.~(\ref{eq:f1})}.

\begin{equation}\label{eq:f1}
F1-score = 2 \times \frac{Precision \times Recall}{Precision + Recall}
\end{equation}


\subsection{Experimental Settings}

In this work, we utilize Pytorch framework and scikit-learn library to build our Fed-LSAE on the hardware configuration of Intel® Xeon® E5-2660 v4 CPU (16 cores - 1.0 GHz), 100 GB RAM and the operating system of Ubuntu 16.04.

The FL-based training process occurs in 10 communication rounds ($R=10$). There are total $n=10$ clients participating in the learning phase where only $k$ agents are selected in each round depending on a fraction factor $C$. In these experiments, $C$ is defined as 1.0, which means $k=C*n=10$ agents in each round.
All participants train their local model in 3 epochs with the batch size of 2048. The loss function is the cross-entropy and the stochastic gradient descent (SGD) optimizer is also used with a learning rate of 0.001 and momentum of 0.9. 

\begin{table}[!b]
\centering
\caption{CNN Architecture}
\label{tab:cnn}
\begin{tabular}{ccccclllll}
\cline{1-5}
\textbf{Layer}     & \textbf{In}  & \textbf{Out} & \textbf{Kernel / Stride / Padding} & \textbf{Activation} &  &  &  &  &  \\ \cline{1-5}
conv1d\_1 & 1   & 64  & 3 x 3 / 1 / 1             & ReLU       &  &  &  &  &  \\
batchnorm1d &  64   &     &   -         & -          &  &  &  &  &  \\
conv1d\_2 & 64  & 128  & 3 x 3 / 1 / 0             & ReLU       &  &  &  &  &  \\
batchnorm1d &  128   &     &     -        & -          &  &  &  &  &  \\
flatten   & -   & - & -                         & -          &  &  &  &  &  \\
fc\_1     & - & 64 & -                         & -          &  &  &  &  &  \\
fc\_2     & 64 & 2  & -                         & -          &  &  &  &  &  \\ \cline{1-5}
\end{tabular}
\end{table}

\begin{table}[!b]
\centering
\caption{LeNet Architecture}
\label{tab:lenet}
\begin{tabular}{ccccclllll}
\cline{1-5}
\textbf{Layer}     & \textbf{In}    & \textbf{Out}   & \textbf{Kernel / Stride / Padding} & \textbf{Activation} &  &  &  &  &  \\ \cline{1-5}
conv2d\_1 & 1     & 64     & 2 x 2 / 1 / 1             & ReLU      &  &  &  &  & \\
batchnorm2d &  64   &     & -            & -          &  &  &  &  &  \\
maxpool2d &       &       & 2 x 2 / 1 / 0             & -         &  &  &  &  & \\
conv2d\_2 & 64     & 128    & 2 x 2 / 1 / 0             & ReLU    &  &  &  &  &   \\
batchnorm2d &  128   &     & -           & -          &  &  &  &  &  \\
maxpool2d &       &       & 2 x 2 / 1 / 0             & -         &  &  &  &  & \\
flatten   & -     & - & -                         & -         &  &  &  &  & \\
fc\_1     & - & 64   & -                         & ReLU      &  &  &  &  & \\
fc\_2     & 64   & 32    & -                         & ReLU      &  &  &  &  & \\
fc\_3     & 32    & 2   & -                         & -          &  &  &  &  &  \\ \cline{1-5}
\end{tabular}
\end{table}

In addition, the ML-based threat detectors are built based on 2 neural network structures named Convolutional Neural Network (CNN) and LeNet, of which architectures are described in \textbf{Table~\ref{tab:cnn}} and \textbf{Table~\ref{tab:lenet}} respectively.

In terms of AE, we use linear layers with bias to build encoder and decoder, as in \textbf{Table~\ref{tab:ae}}. Each PLR of benign models is respectively fed into AE model to train in $e=20$ epochs with an Adam optimizer and the learning rate of 0.001. The input and output dimension used in AE are the same, which represent the number of features of each PLR vector.

\begin{table}[!b]
\centering
\caption{Structure of Encoder and Decoder in AE architecture}
\label{tab:ae}
\begin{tabular}{cccc}
\hline
\textbf{Layer} & \textbf{Input} & \textbf{Output} & \textbf{Activation} \\ \hline
\multicolumn{4}{l}{\textbf{Encoder}}                                    \\ \hline
Linear         & input\_dim$^1$             & 512             & ReLU                \\
Linear         & 512            & 128             & ReLU                \\
Linear         & 128            & 64              & ReLU                \\
Linear         & 64             & 16              & -                   \\ \hline
\multicolumn{4}{l}{\textbf{Decoder}}                                    \\ \hline
Linear         & 16             & 64              & ReLU                \\
Linear         & 64             & 128             & ReLU                \\
Linear         & 128            & 512             & ReLU                \\
Linear         & 512            & output\_dim$^1$              & Tanh               \\ \hline
\multicolumn{4}{l}{\small $^1$ Dimension of input and output respectively} \\
\end{tabular}
\end{table}

In GAN-based poisoning attacks, we implement a GAN architecture with the hyperparameters of epochs = 20, batch\_size = 512, and the Adam optimizer with a learning rate of 0.0001. The generator $G$ and discriminator $D$ are designed with 5 layers, with the detailed structure in \textbf{Table~\ref{tab:idsgan}}.

\begin{table}[!t]
\centering
\caption{Structure of Generator $G$ and Discriminator $D$ in IDSGAN architecture}
\label{tab:idsgan}
\begin{tabular}{lccc}
\hline
\textbf{Layer} & \textbf{Input} & \textbf{Output} & \textbf{Activation} \\ \hline
\multicolumn{4}{l}{\textbf{Generator}}                                                  \\ \hline
Linear             & input\_dim$^2$     & input\_dim//2    & ReLU                   \\
Linear             & input\_dim//2   & input\_dim//2    & ReLU                   \\
Linear             & input\_dim//2   & input\_dim//2    & ReLU                   \\
Linear             & input\_dim//2   & input\_dim//2   & ReLU                   \\
Linear             & input\_dim//2  & output\_dim$^2$     & -                       \\ \hline
\multicolumn{4}{l}{\textbf{Discriminator}}                                              \\ \hline
Linear             & input\_dim     & input\_dim*2    & LeakyReLU              \\
Linear             & input\_dim*2   & input\_dim*2    & LeakyReLU              \\
Linear             & input\_dim*2   & input\_dim*2    & LeakyReLU              \\
Linear             & input\_dim*2   & input\_dim//2   & LeakyReLU              \\
Linear             & input\_dim//2  & output\_dim     & -                  \\ \hline
\multicolumn{4}{l}{\small $^2$ Dimension of input and output respectively} \\
\end{tabular}
\end{table}

\begin{figure*}[!b]
 \centering
 \begin{tabular}{cc}
 \subfloat[CIC-ToN-IoT]{%
 \includegraphics[width=0.4\textwidth]{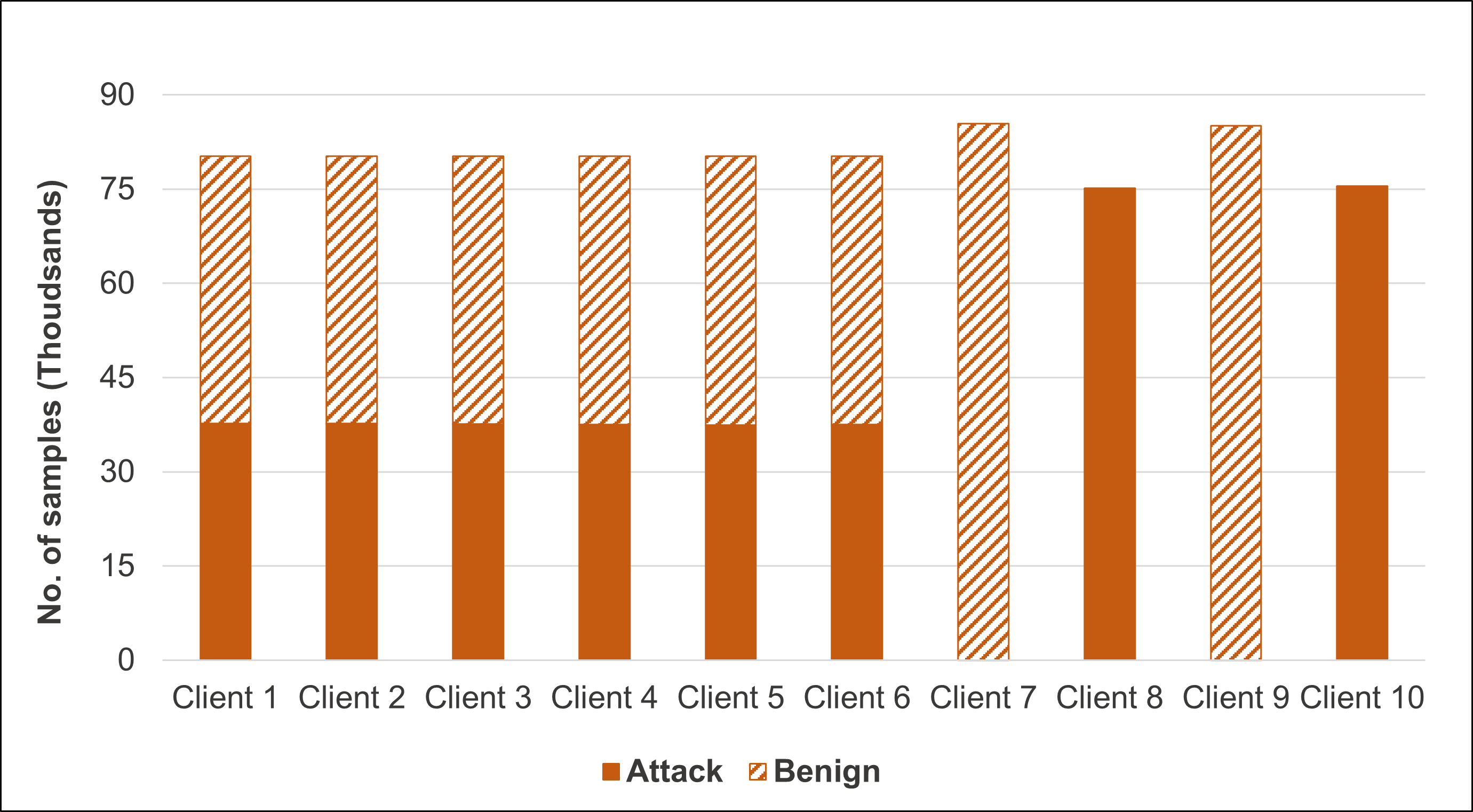}%
 } \hfill{}
 \subfloat[N-BaIoT]{%
 \includegraphics[width=0.4\textwidth]{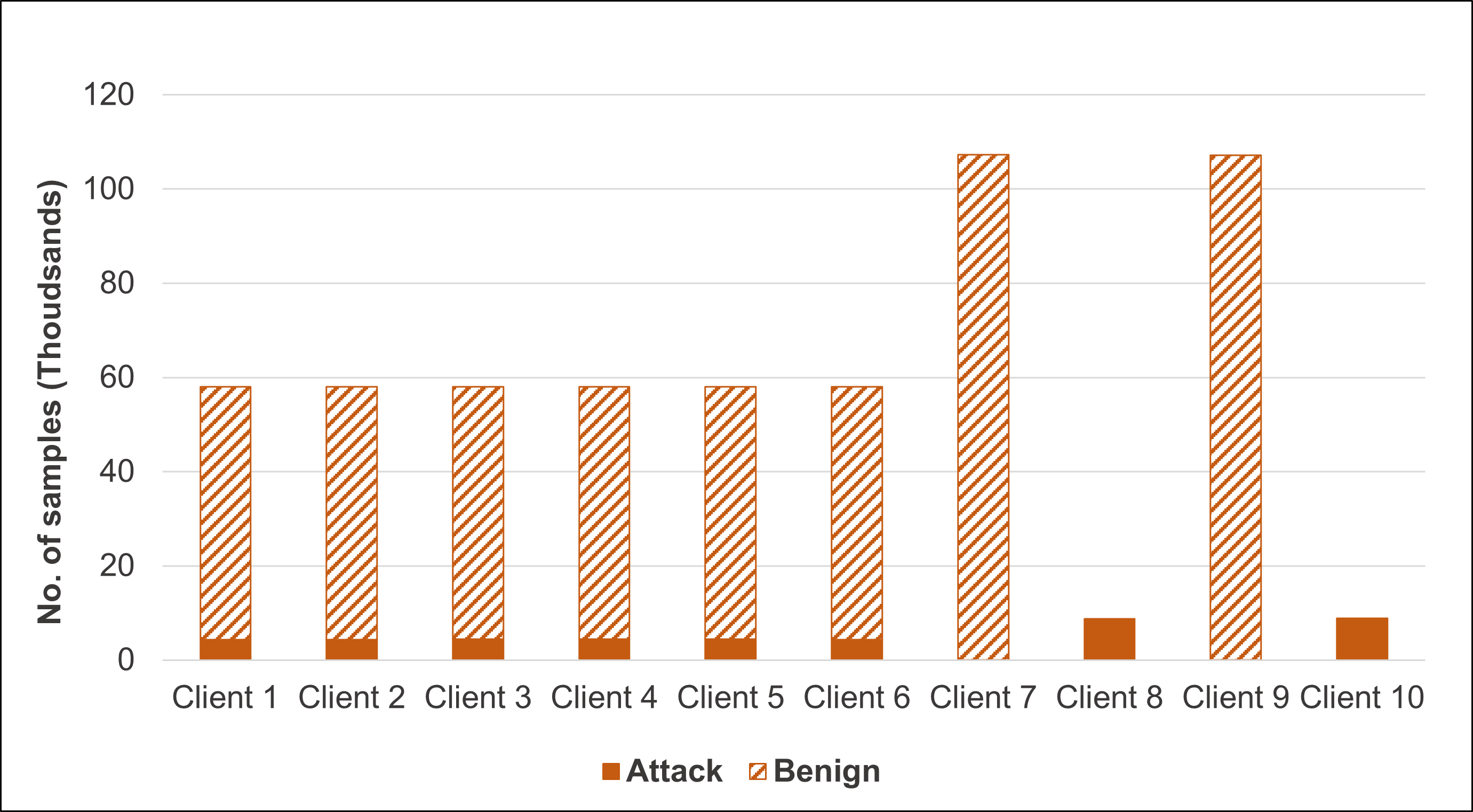}%
 }
 \end{tabular}
 \caption{The data distribution among clients on \textit{(a)} CIC-ToN-IoT and \textit{(b)} N-BaIoT datasets in non-IID cases.}
 \label{fig: s3-nonIID}
\end{figure*}
Note that, all following experiments are performed 5 times, and overall results are the average to ensure accuracy and reliability of our findings.

\subsection{Experimental Scenarios}

\subsubsection{Scenario 1 - Baseline performance of federated threat detectors}
This scenario aims to evaluate the baseline effectiveness of two ML-based threat detector models, including CNN and LeNet, on CIC-ToN-IoT and N-BaIoT datasets in the context of FL. In other words, only benign clients participate in training the FL-based model, whose aggregation is based on the FedAvg algorithm as in \textbf{Eq.~(\ref{eq:fedavg})}. Moreover, each local agent has the same data distribution as the others in terms of the number of training samples and the ratio of benign and malicious labels. Our goal is to build different FL-based threat detectors that have the ability to detect abnormal traffic in IoT networks.

\subsubsection{Scenario 2 - Evaluation on the performance of Fed-LSAE in eliminating poisoned updates} \label{Scenario_2}
To assess the effectiveness of our proposed defense framework, we clarify the robustness of FL-based threat detectors against poisoning attacks after integrating the Fed-LSAE module in IID environment. For more details, 4 out of 10 clients are assumed as compromised agents (adversaries) to conduct 3 typical strategies of poisoning attacks, as described in \textbf{Section \ref{subsec: attack_strategy}}, throughout the FL-based learning phase. In weight-scaling model poisoning attacks, adversaries try to scale their poisoned weight parameters up to 10 times. In this scenario, we observe the negative impact of these attacks on the overall performance of FL-based detector models and the usefulness of our Fed-LSAE in defeating adversaries to maintain the stability of these models.

\subsubsection{Scenario 3 - Comparison of defense performance to other methods}
This scenario reveals the outstanding features of our Fed-LSAE compared to the previous proposed FedCC scheme \cite{Jeong2022_FedCC}. To have a reliable comparison, this evaluation is performed in the same context of Median \cite{fang2021_ByzantinePA} poisoning attack, an untargeted model poisoning attack as in experiments of FedCC \cite{Jeong2022_FedCC}. Besides, to ensure the objectivity of our approach, we conduct this experiment on two models, including CNN and LeNet, following a similar methodology as that employed in the FedCC study. All the metrics and CKA scores are averaged to observe the stability of each scheme when dealing with this attack.
Our desired objective is to demonstrate how Fed-LSAE outperforms FedCC in the 3 following aspects:

\begin{itemize}
    \item The stability in dealing with Median attacks in the case of IID data.
    \item The ability to detect poisonous agents.
    \item The performance when the rest of the benign clients follow the non-IID pattern, which is depicted in \textbf{Fig. \ref{fig: s3-nonIID}}. More specifically, on both datasets, the first six clients, including four adversaries, have the same data distribution whereas the remaining four clients will follow other patterns. Clients 7 and 9 contain 100\% benign samples, while clients 8 and 10 collect only attack data traffic. 
\end{itemize}


\subsection{Experimental Results}

\subsubsection{Scenario 1}
\begin{figure*}[!t]
 \centering
 \begin{tabular}{cc}
 \subfloat[CNN]{%
 \includegraphics[width=0.24\textwidth]{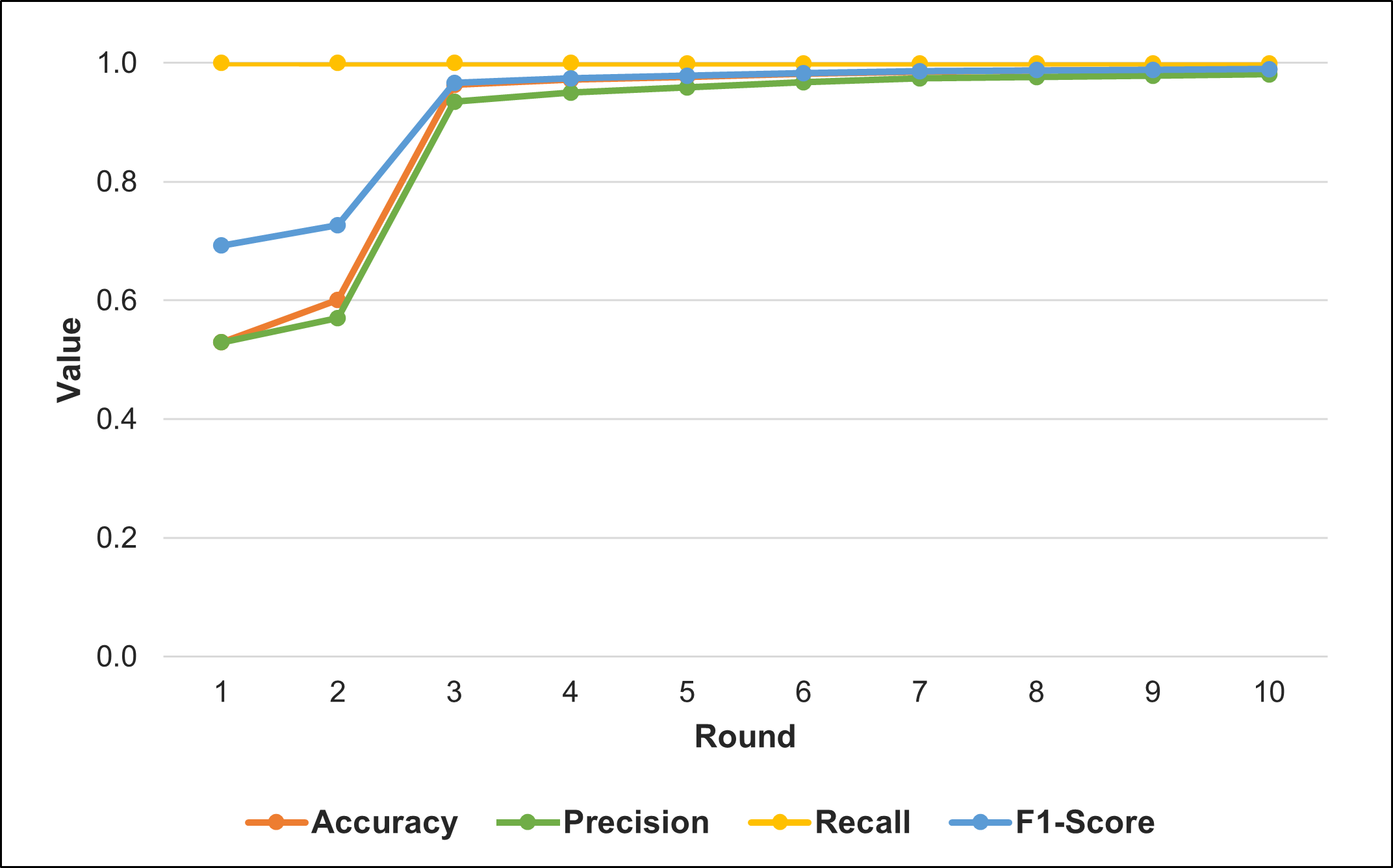}%
 } \hfill{}
 \subfloat[LeNet]{%
 \includegraphics[width=0.24\textwidth]{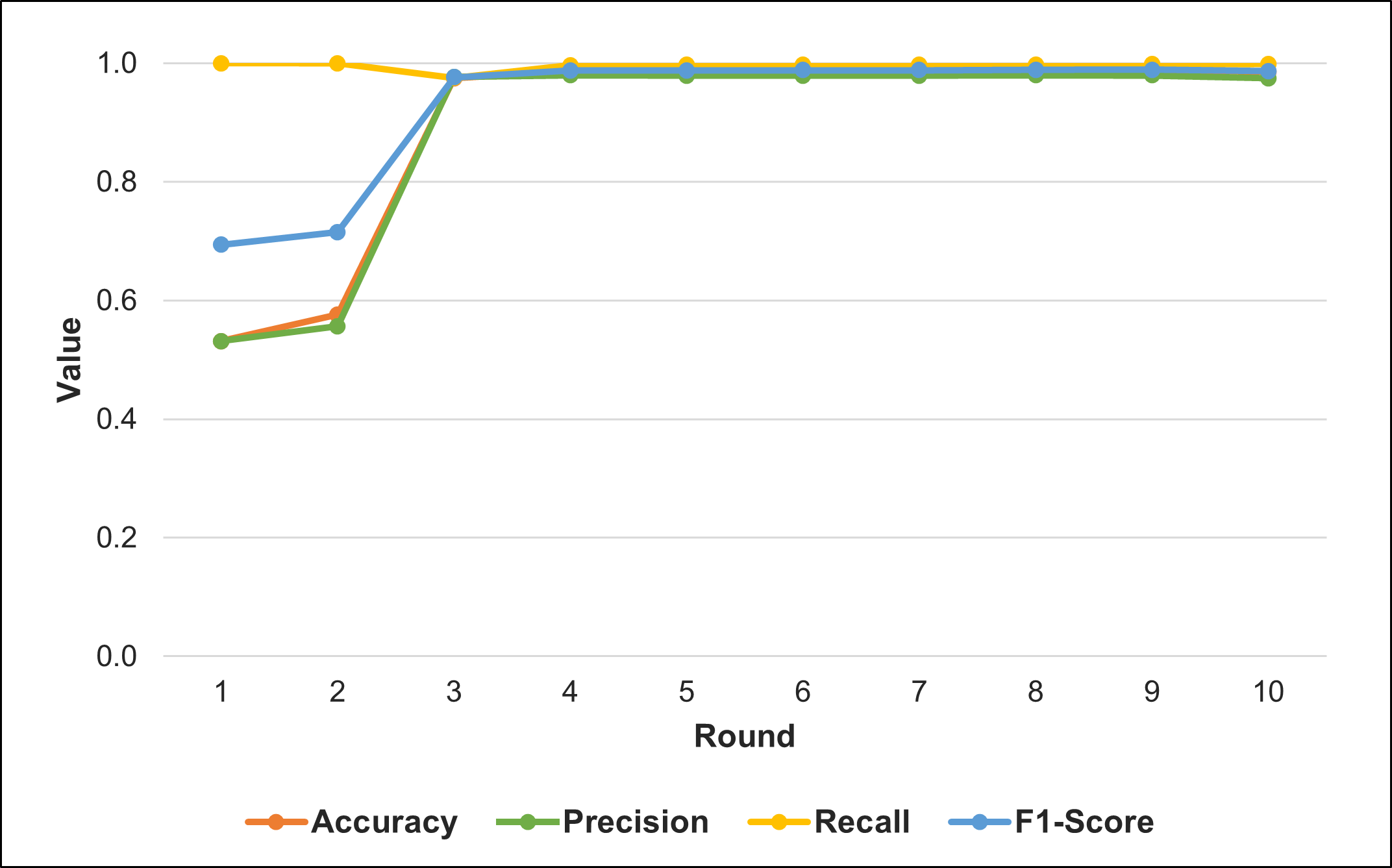}%
 }\hfill{}
 \subfloat[CNN]{%
 \includegraphics[width=0.24\textwidth]{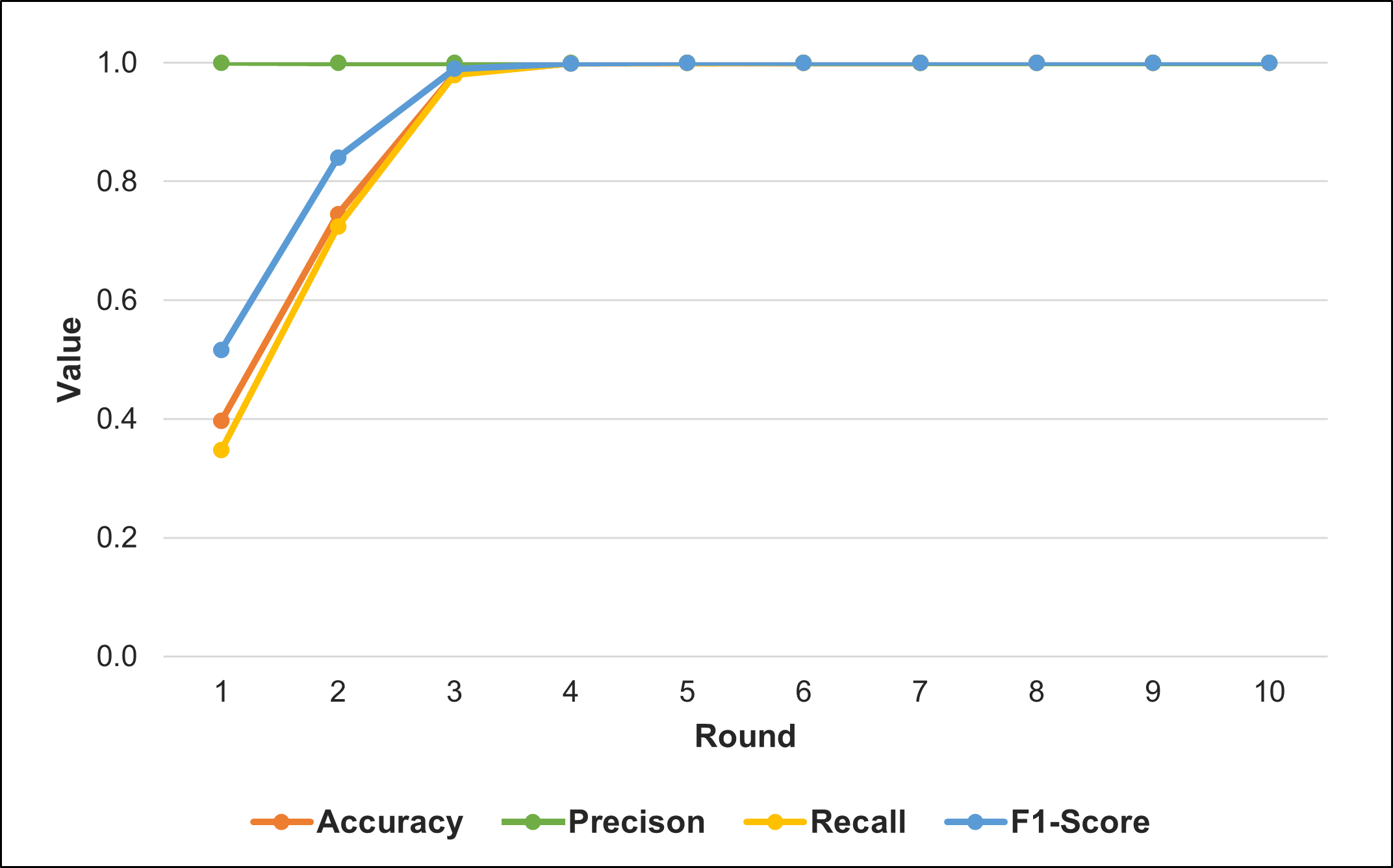}%
 } \hfill{}
 \subfloat[LeNet]{%
 \includegraphics[width=0.24\textwidth]{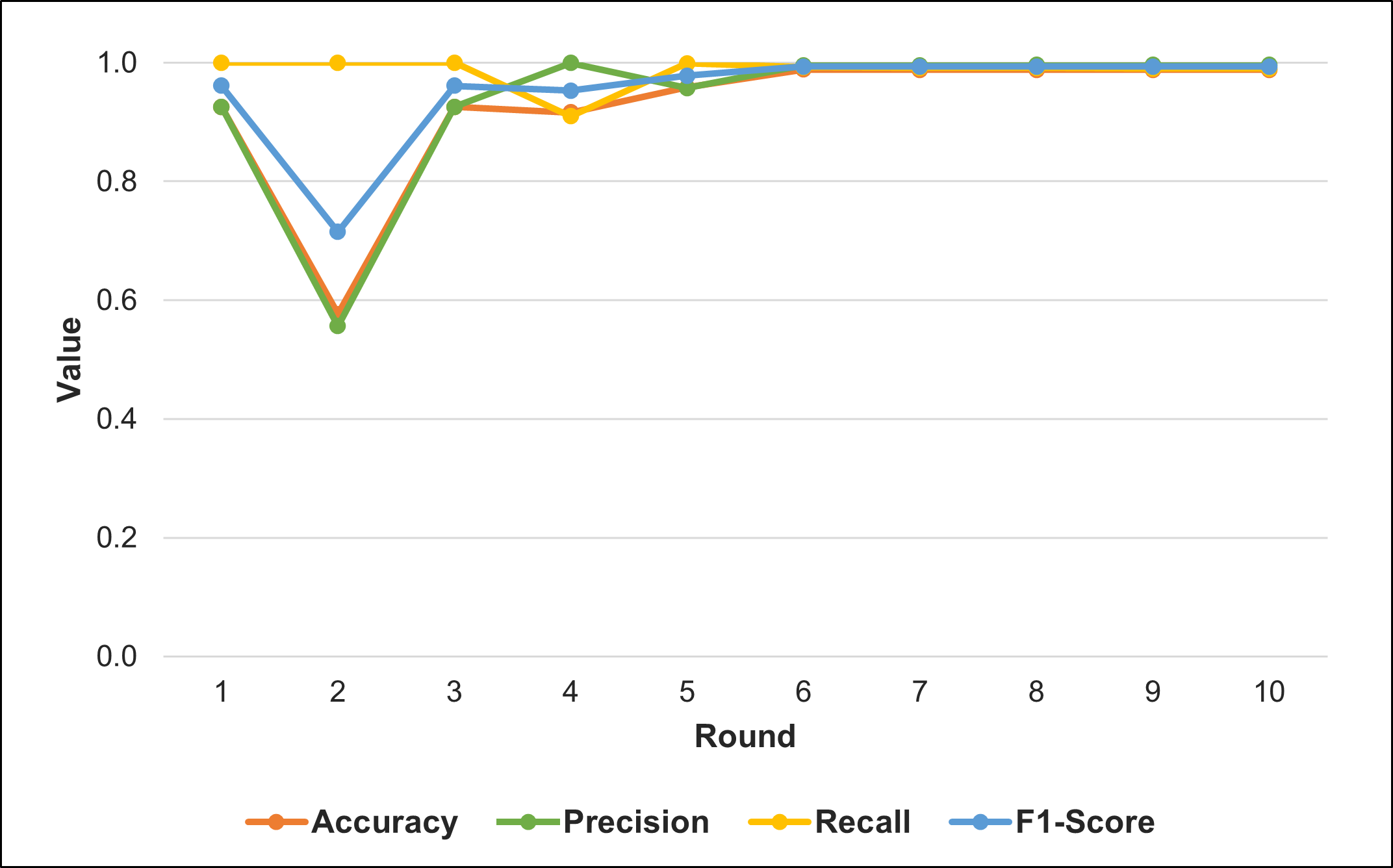}%
 }
 \end{tabular}
 \caption{FL-based training process of threat detector on \textit{(a,b)} CIC-ToN-IoT and \textit{(c,d)} N-BaIoT datasets.}
 \label{fig: s1-training}
\end{figure*}

The performance of two FL-based threat detector models is shown in \textbf{Fig.~\ref{fig: s1-training}}, in terms of the Accuracy, Precision, Recall and F1-score. Although LeNet model has witnessed a fluctuation during the first three rounds on N-BaIoT dataset, the performance of the model on both datasets has rapidly grown to the convergence point and achieved more than 98\% in all metrics. These results prove the effectiveness of these FL models in detecting cyber threats in the context of IoT networks.

\begin{figure*}[!t]
 \centering
 \def\twidth{0.4}
 \subfloat[CNN]{%
 \includegraphics[width=0.24\textwidth]{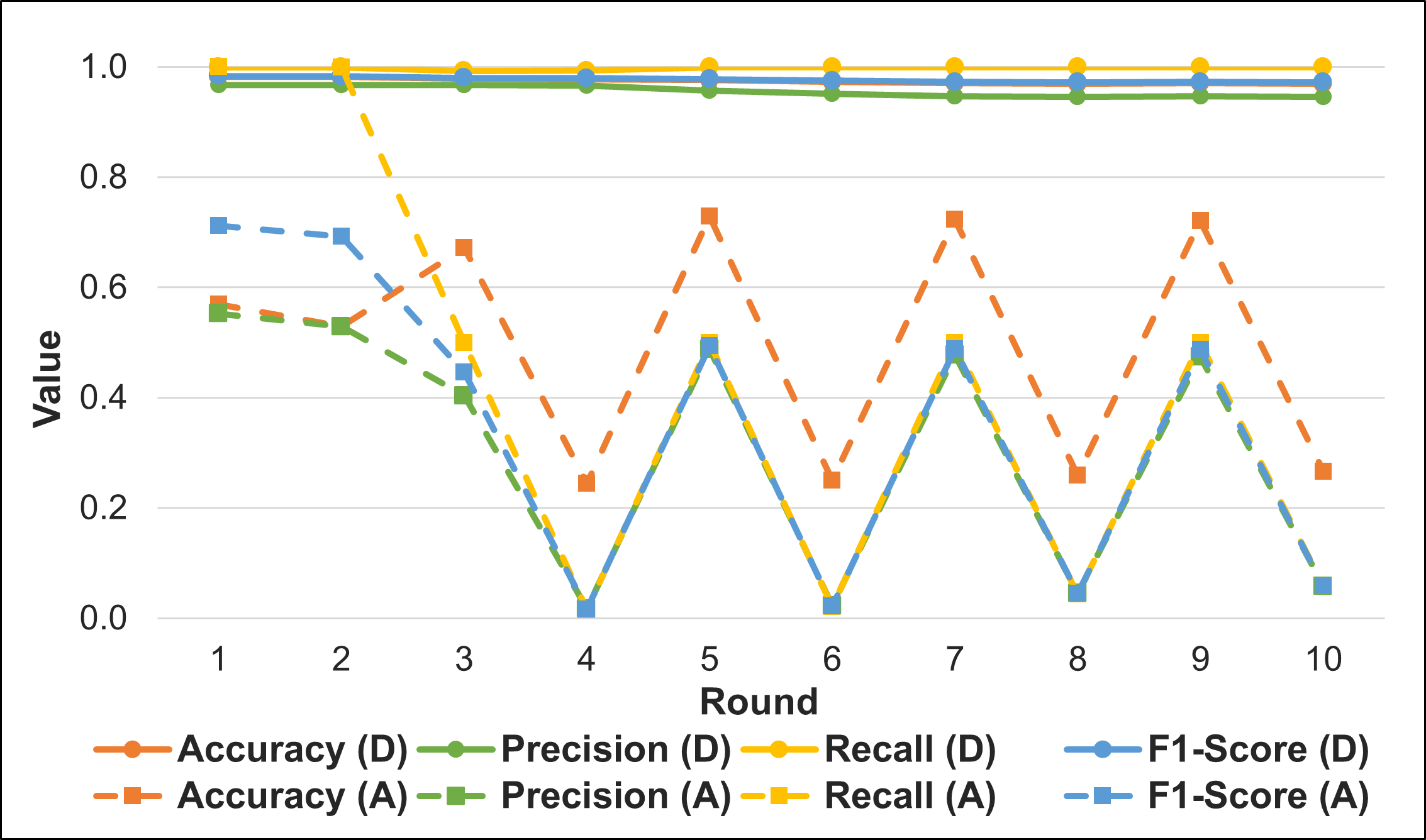}%
 } \hfill
 \subfloat[LeNet]{%
 \includegraphics[width=0.24\textwidth]{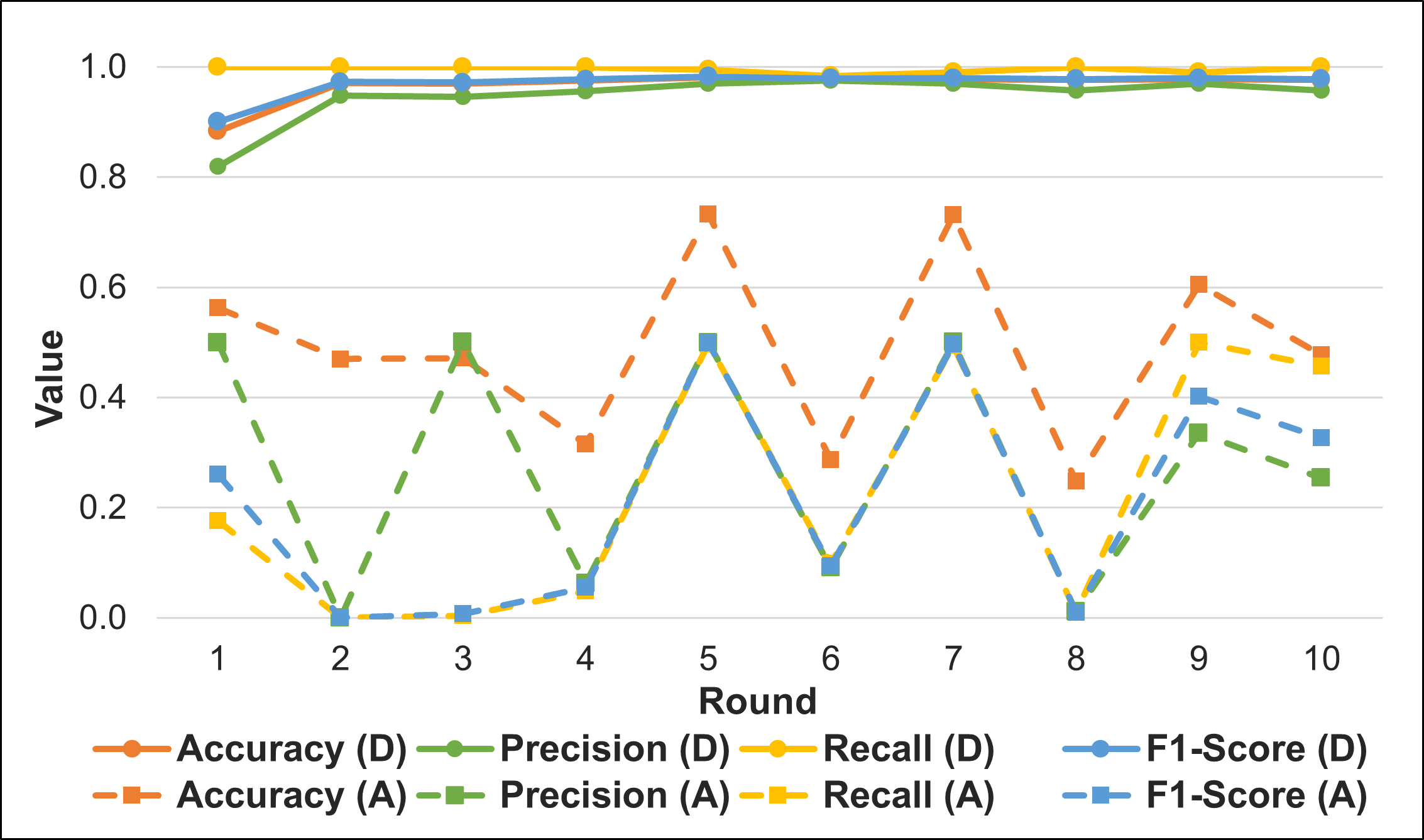}%
 } \hfill
 \subfloat[CNN]{%
 \includegraphics[width=0.24\textwidth]{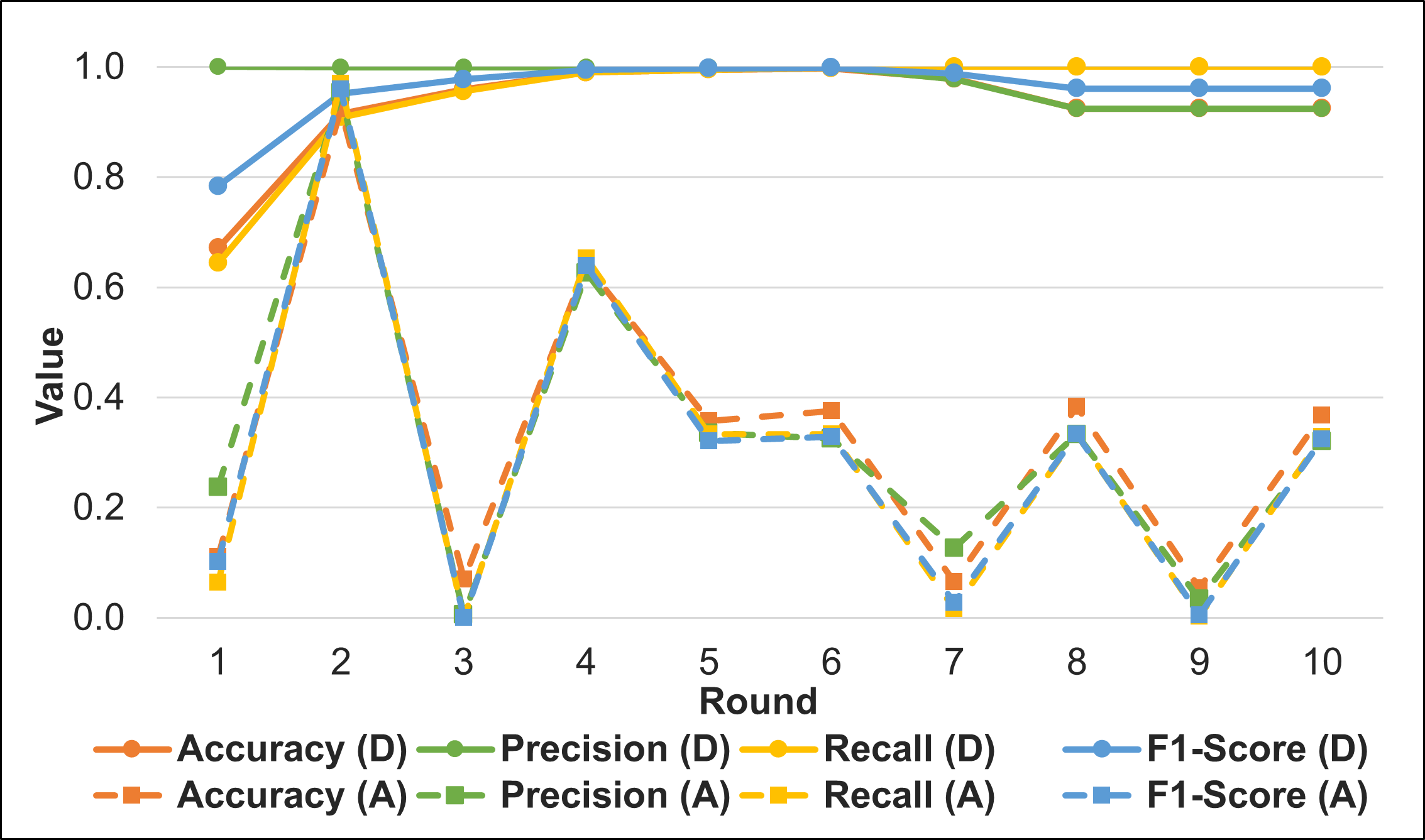}%
 } \hfill
 \subfloat[LeNet]{%
 \includegraphics[width=0.24\textwidth]{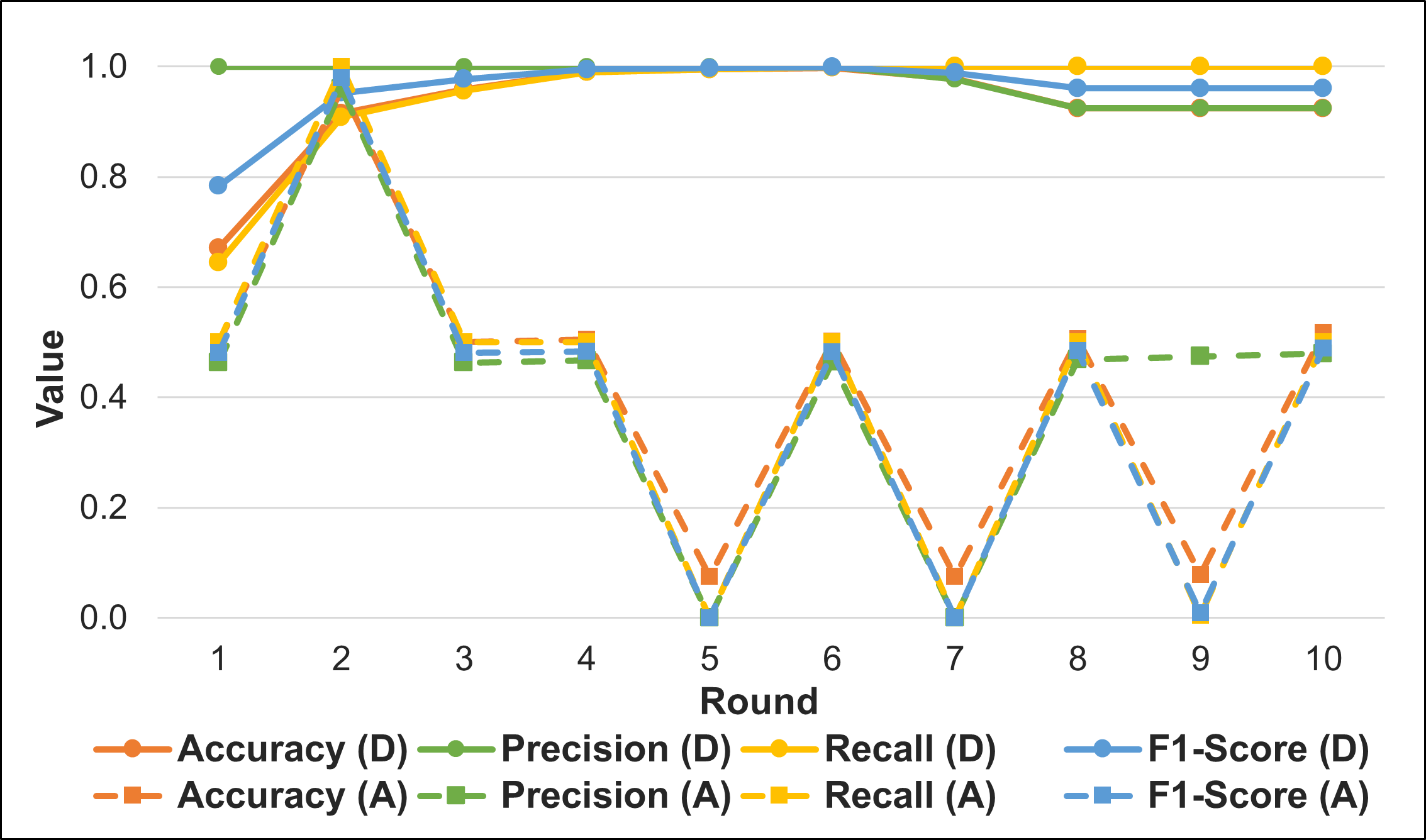}%
 } \hfill
 
 \caption{Fed-LSAE performance against Label Flipping attacks on \textit{(a,b)} CIC-ToN-IoT and \textit{(c,d)} N-BaIoT datasets.}
 \label{fig: s2-defense-lf}
\end{figure*}

\begin{figure*}[!t]
 \centering
 \def\twidth{0.4}
 \subfloat[CNN]{%
 \includegraphics[width=0.24\textwidth]{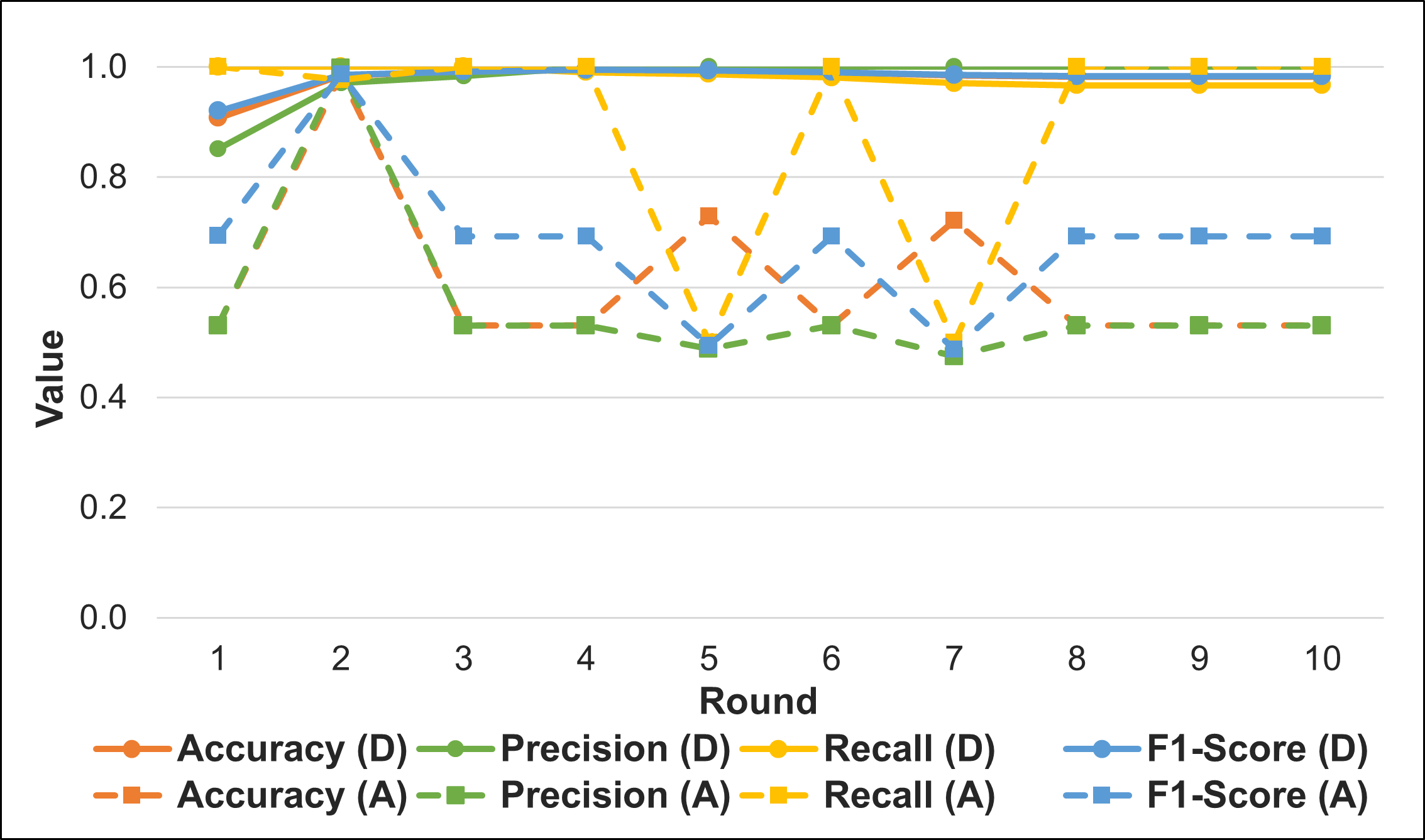}%
 } \hfill
 \subfloat[LeNet]{%
 \includegraphics[width=0.24\textwidth]{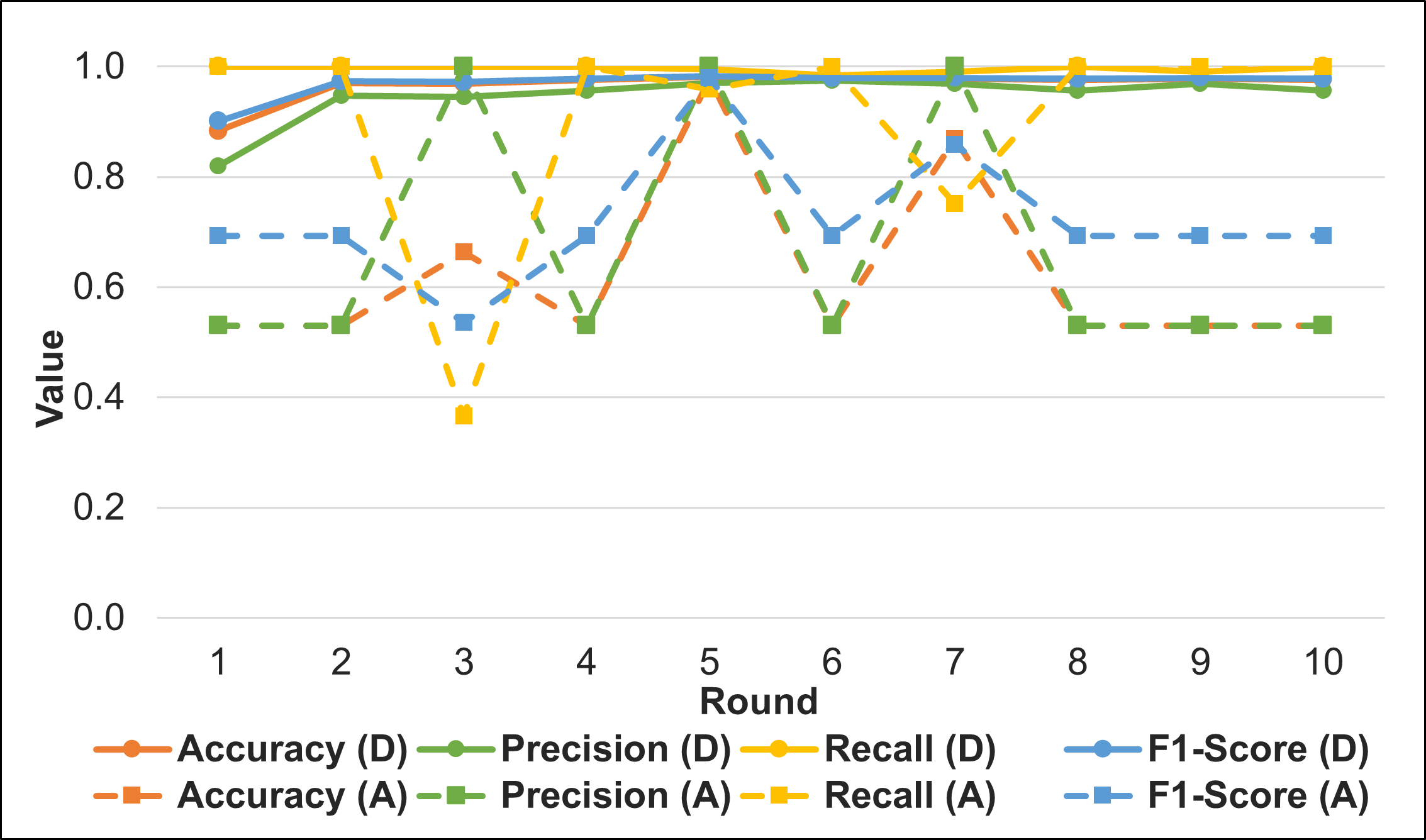}%
 } \hfill
 \subfloat[CNN]{%
 \includegraphics[width=0.24\textwidth]{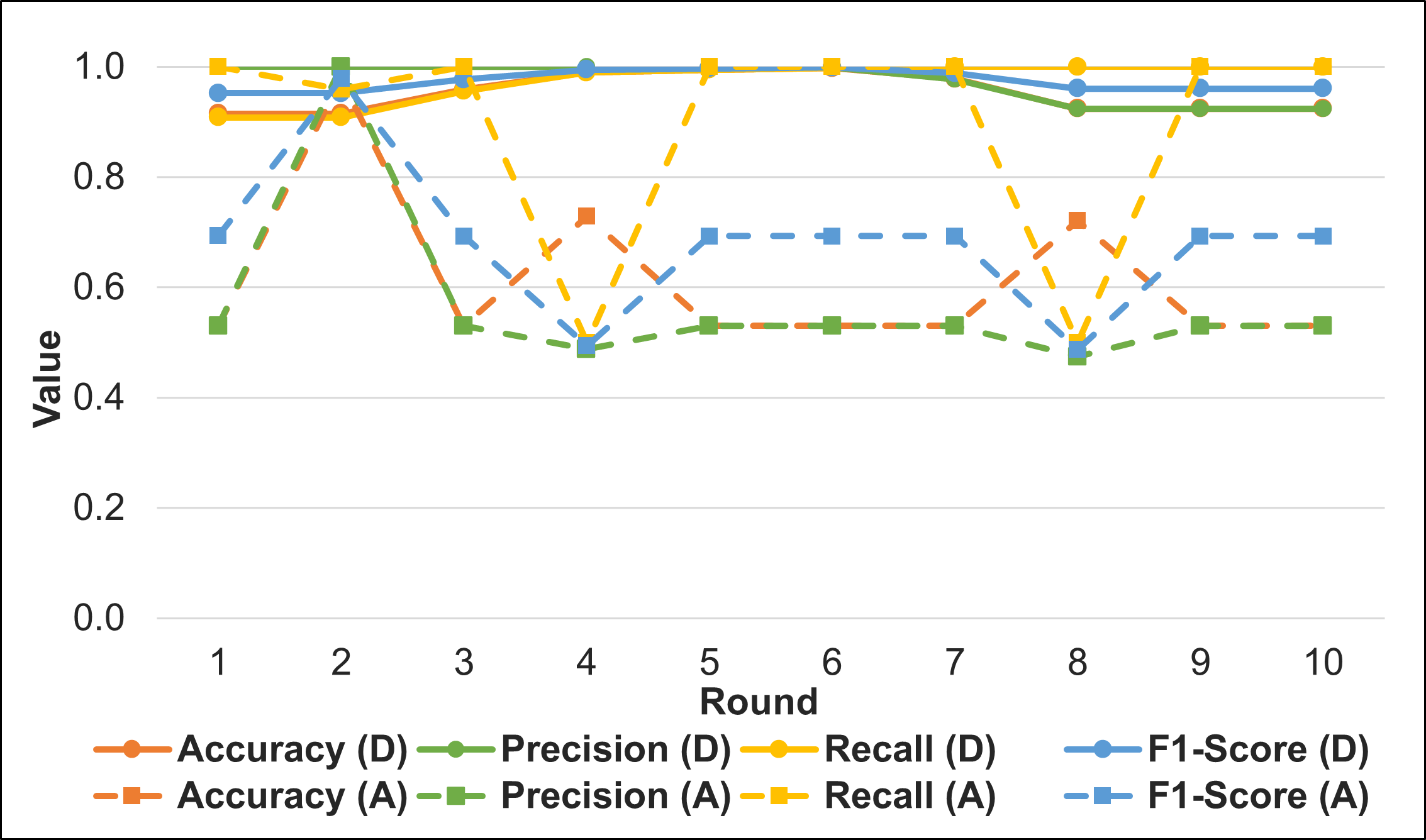}%
 } \hfill
 \subfloat[LeNet]{%
 \includegraphics[width=0.24\textwidth]{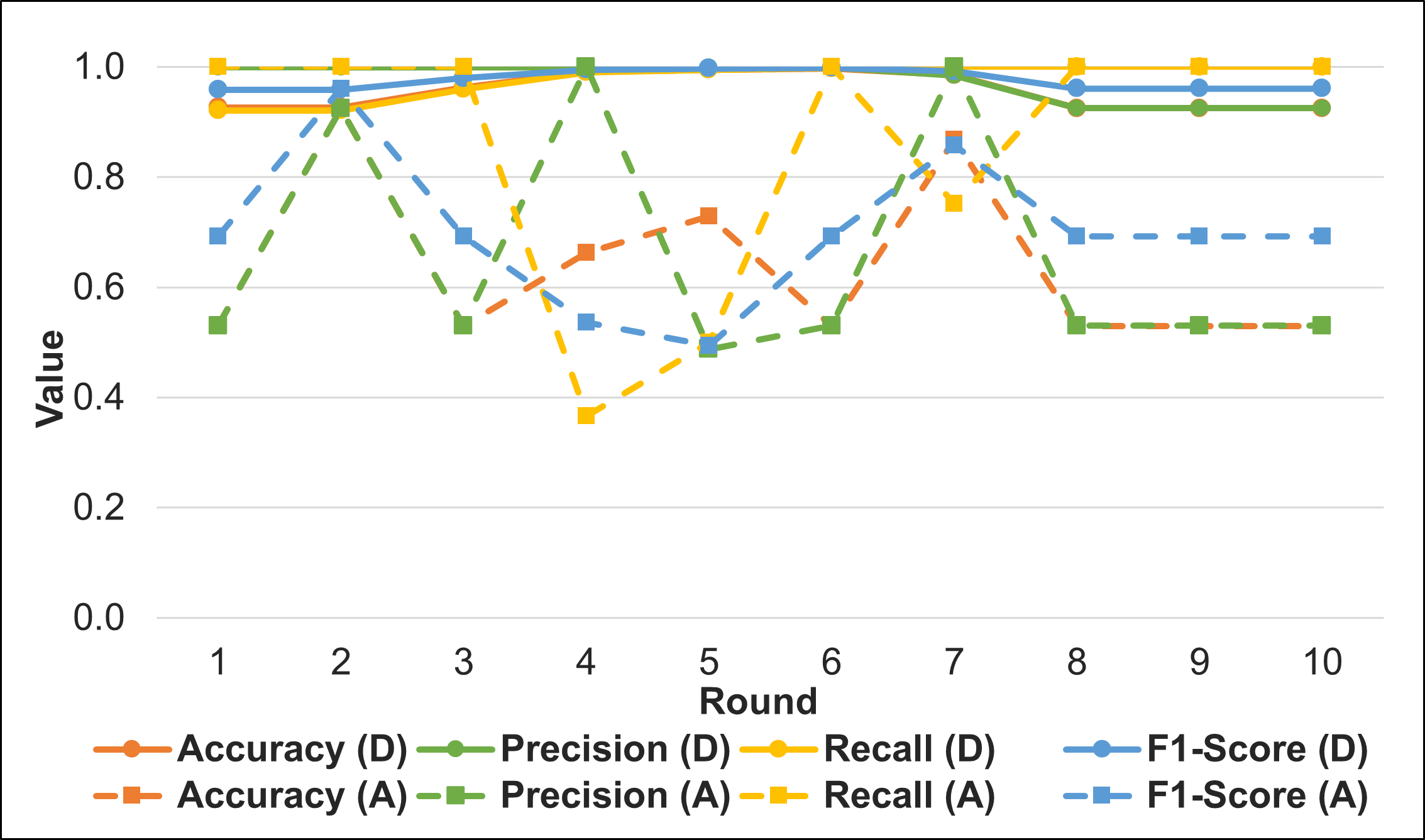}%
 } \hfill
 
 \caption{Fed-LSAE performance against GAN-based attacks on \textit{(a,b)} CIC-ToN-IoT and \textit{(c,d)} N-BaIoT datasets.}
 \label{fig: s2-defense-gan}
\end{figure*}

\begin{figure*}[!t]
 \centering
 \def\twidth{0.4}
 \subfloat[CNN]{%
 \includegraphics[width=0.24\textwidth]{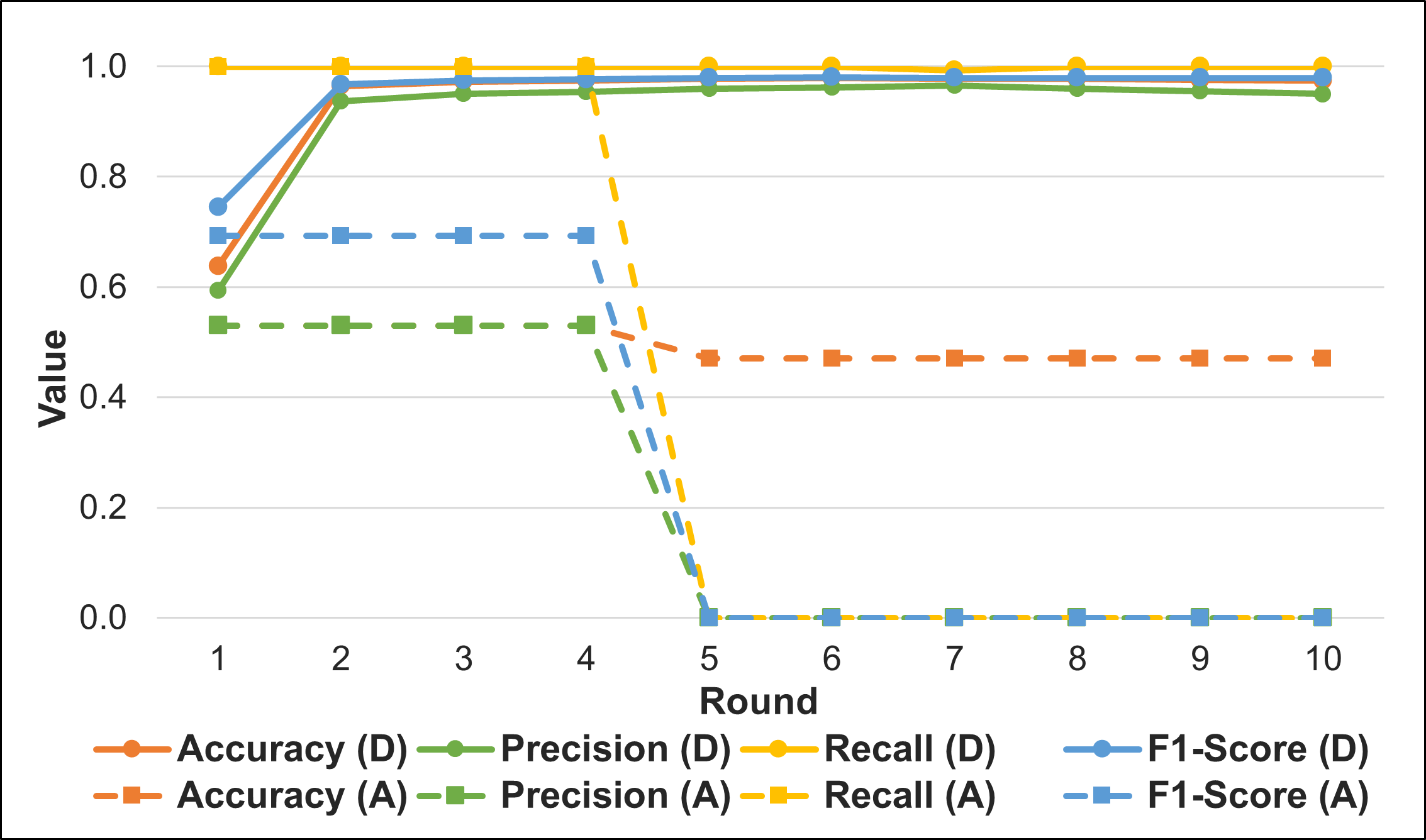}%
 } \hfill
 \subfloat[LeNet]{%
 \includegraphics[width=0.24\textwidth]{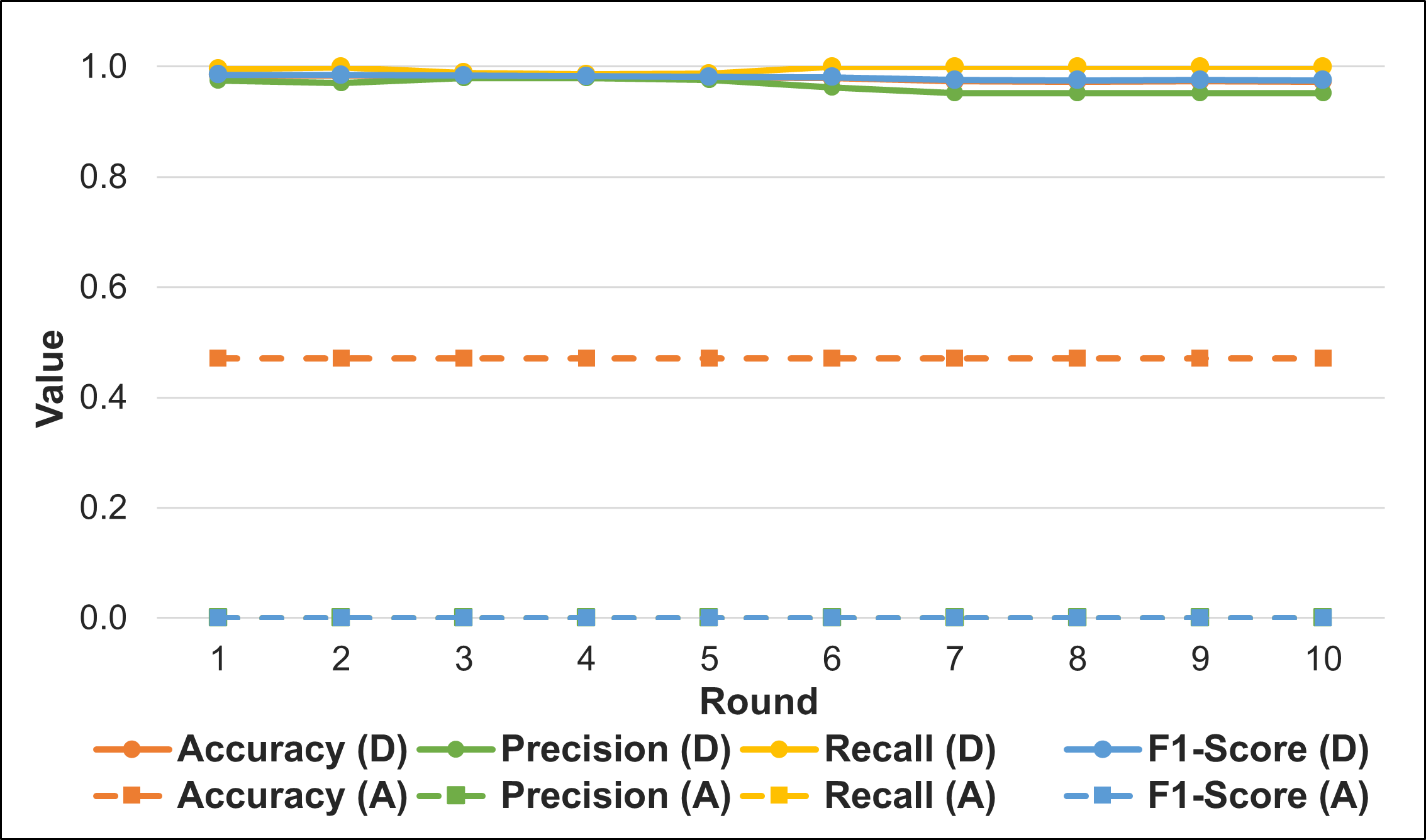}%
 } \hfill
 \subfloat[CNN]{%
 \includegraphics[width=0.24\textwidth]{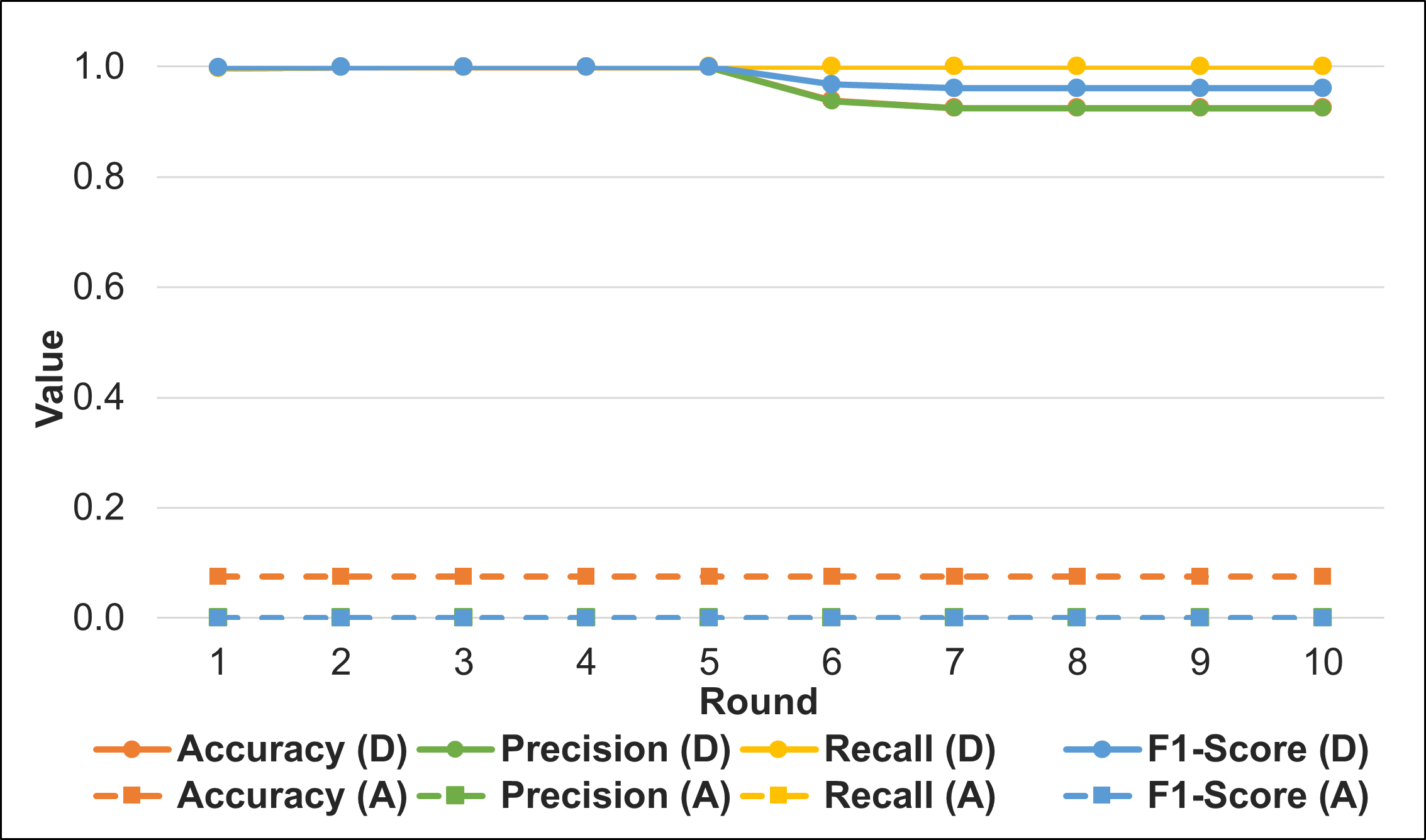}%
 } \hfill
 \subfloat[LeNet]{%
 \includegraphics[width=0.24\textwidth]{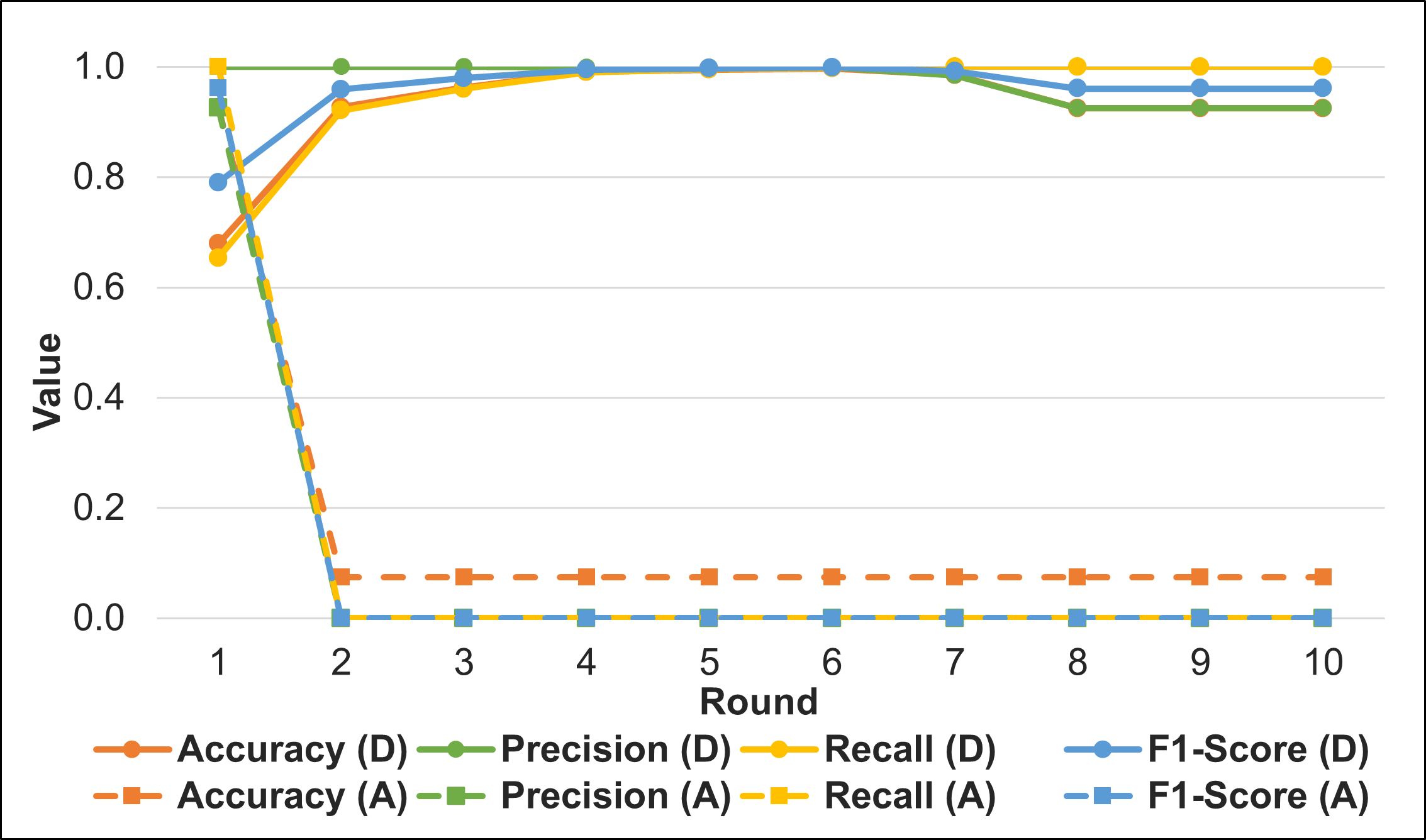}%
 } \hfill
 
 \caption{Fed-LSAE performance against Model Poisoning attacks on \textit{(a,b)} CIC-ToN-IoT and \textit{(c,d)} N-BaIoT datasets.}
 \label{fig: s2-defense-mp}
\end{figure*}

\subsubsection{Scenario 2}
The detailed results of our proposed method in eliminating poisoned updates are summarized in \textbf{Fig.~\ref{fig: s2-defense-lf}}, \textbf{Fig.~\ref{fig: s2-defense-gan}} and \textbf{Fig.~\ref{fig: s2-defense-mp}}. Thereby, $A$ and dot lines describe metrics in case of attack without Fed-LSAE, while $D$ and solid lines indicate results with defense by Fed-LSAE.  In all three poisoning attack strategies, there has been a sharp decline in the performance of both FL-based threat detectors without defense. In Label Flipping (\textbf{Fig.~\ref{fig: s2-defense-lf}}) and GAN-based attacks (\textbf{Fig.~\ref{fig: s2-defense-gan}}), the detecting rate of FL-based models has fluctuated around roughly 50\% across all metrics. On the other hand, the models seem to be completely damaged by the weight-scaling model poisoning method when Precision, Recall and F1-Score benchmarks reach exactly 0\% in almost all communication rounds respectively. It implies that weight-scaling model poisoning is easier to conduct and more efficient than the others, since it directly affects the global model parameters. 

In the context of thwarting poisoned updates, the performance of our Fed-LSAE has rapidly reached the convergence point since $3^{rd}$ communication round and achieved over 98\% across all four metrics. By learning data representations via PLR vector, which is the most distinguishable layer out of all layers in a neural network, our Fed-LSAE has proved its effectiveness in eliminating both data poisoning and model poisoning attacks from ruining robust FL-based detection systems.

\subsubsection{Scenario 3}
The descriptive statistics in \textbf{Table~\ref{tab:fedcc-fedlsae-fang-IID}} have revealed that our Fed-LSAE achieves a better detection rate than FedCC in terms of Median attacks in all cases. As we can see, the Fed-LSAE framework can recognize all 4 adversaries out of 10 clients and witnessed a stable trend in its defensive performance during 10 rounds, with an average of over 93\% in Accuracy and 96\% in F1-Score across four cases. Meanwhile, the FedCC scheme still experienced difficulties in detecting Median attacks with average results of approximately 66\% and 77\% in Accuracy and F1-Score respectively in the worst case of LeNet model on N-BaIoT dataset. 

\begin{table*}[!t]
\centering
\caption{Defense performance comparison between FedCC and Fed-LSAE when tackling with Median attacks in IID case}
\begin{tabular}{|c|c|cccc|cccc|}
\hline
\multirow{2}{*}{\textbf{Model}} & \multirow{2}{*}{\textbf{Scheme}} & \multicolumn{4}{c|}{\textbf{CIC-ToN-IoT}}                                                                                                & \multicolumn{4}{c|}{\textbf{N-BaIoT}}                                                                                                    \\ \cline{3-10} 
                                &                                  & \multicolumn{1}{c|}{Accuracy}         & \multicolumn{1}{c|}{Precision}        & \multicolumn{1}{c|}{Recall}           & F1-Score         & \multicolumn{1}{c|}{Accuracy}         & \multicolumn{1}{c|}{Precision}        & \multicolumn{1}{c|}{Recall}           & F1-Score         \\ \hline
\multirow{2}{*}{\textbf{CNN}}   & \textbf{FedCC}                   & \multicolumn{1}{c|}{0.98923}          & \multicolumn{1}{c|}{0.9932}           & \multicolumn{1}{c|}{\textbf{0.98676}} & 0.98981          & \multicolumn{1}{c|}{0.94178}          & \multicolumn{1}{c|}{0.94172}          & \multicolumn{1}{c|}{0.99995}          & 0.96973          \\ \cline{2-10} 
                                & \textbf{Fed-LSAE}                & \multicolumn{1}{c|}{\textbf{0.99118}} & \multicolumn{1}{c|}{\textbf{0.99923}} & \multicolumn{1}{c|}{0.98413} & \textbf{0.99159} & \multicolumn{1}{c|}{\textbf{0.95613}} & \multicolumn{1}{c|}{\textbf{0.95623}} & \multicolumn{1}{c|}{\textbf{0.99978}} & \textbf{0.97719} \\ \hline
\multirow{2}{*}{\textbf{LeNet}} & \textbf{FedCC}                   & \multicolumn{1}{c|}{0.80766}          & \multicolumn{1}{c|}{0.80428}          & \multicolumn{1}{c|}{\textbf{0.99928}} & 0.87304          & \multicolumn{1}{c|}{0.65797}          & \multicolumn{1}{c|}{0.99999}          & \multicolumn{1}{c|}{0.63044}          & 0.76556          \\ \cline{2-10} 
                                & \textbf{Fed-LSAE}                & \multicolumn{1}{c|}{\textbf{0.96116}} & \multicolumn{1}{c|}{\textbf{0.94727}} & \multicolumn{1}{c|}{0.99602}          & \textbf{0.96819} & \multicolumn{1}{c|}{\textbf{0.93106}} & \multicolumn{1}{c|}{\textbf{0.93251}} & \multicolumn{1}{c|}{\textbf{0.99841}} & \textbf{0.96420}  \\ \hline
\end{tabular}
\label{tab:fedcc-fedlsae-fang-IID}
\end{table*}

\begin{table*}[!b]
\centering
\caption{Defense performance comparison between FedCC and Fed-LSAE when tackling with Median attacks in non-IID case}
\begin{tabular}{|c|c|cccc|cccc|}
\hline
\multirow{2}{*}{\textbf{Model}} & \multirow{2}{*}{\textbf{Scheme}} & \multicolumn{4}{c|}{\textbf{CIC-ToN-IoT}}                                                                                                & \multicolumn{4}{c|}{\textbf{N-BaIoT}}                                                                                                    \\ \cline{3-10} 
                                &                                  & \multicolumn{1}{c|}{Accuracy}         & \multicolumn{1}{c|}{Precision}        & \multicolumn{1}{c|}{Recall}           & F1-Score         & \multicolumn{1}{c|}{Accuracy}         & \multicolumn{1}{c|}{Precision}        & \multicolumn{1}{c|}{Recall}           & F1-Score         \\ \hline
\multirow{2}{*}{\textbf{CNN}}   & \textbf{FedCC}                   & \multicolumn{1}{c|}{0.54266}          & \multicolumn{1}{c|}{0.34839}          & \multicolumn{1}{c|}{0.6}              & 0.43725          & \multicolumn{1}{c|}{0.6816}           & \multicolumn{1}{c|}{0.63453}          & \multicolumn{1}{c|}{\textbf{0.9}}     & 0.7443          \\ \cline{2-10} 
                                & \textbf{Fed-LSAE}                & \multicolumn{1}{c|}{\textbf{0.60364}} & \multicolumn{1}{c|}{\textbf{0.5928}}  & \multicolumn{1}{c|}{\textbf{0.9988}}  & \textbf{0.73679} & \multicolumn{1}{c|}{\textbf{0.72094}} & \multicolumn{1}{c|}{\textbf{0.67387}} & \multicolumn{1}{c|}{\textbf{0.9}}     & \textbf{0.77069} \\ \hline
\multirow{2}{*}{\textbf{LeNet}} & \textbf{FedCC}                   & \multicolumn{1}{c|}{0.59378}          & \multicolumn{1}{c|}{0.57047}          & \multicolumn{1}{c|}{0.85507}          & 0.65917          & \multicolumn{1}{c|}{0.63907}          & \multicolumn{1}{c|}{0.60248}          & \multicolumn{1}{c|}{0.88016}          & 0.69647          \\ \cline{2-10} 
                                & \textbf{Fed-LSAE}                & \multicolumn{1}{c|}{\textbf{0.71483}} & \multicolumn{1}{c|}{\textbf{0.71731}} & \multicolumn{1}{c|}{\textbf{0.99333}} & \textbf{0.81232} & \multicolumn{1}{c|}{\textbf{0.72651}} & \multicolumn{1}{c|}{\textbf{0.73692}} & \multicolumn{1}{c|}{\textbf{0.98014}} & \textbf{0.82102} \\ \hline
\end{tabular}
\label{tab: s3-nonIID-fang}
\end{table*}

\begin{figure}[!t]
 \centering
 \def\twidth{0.4}
 \subfloat[CNN]{%
 \includegraphics[width=0.24\textwidth]{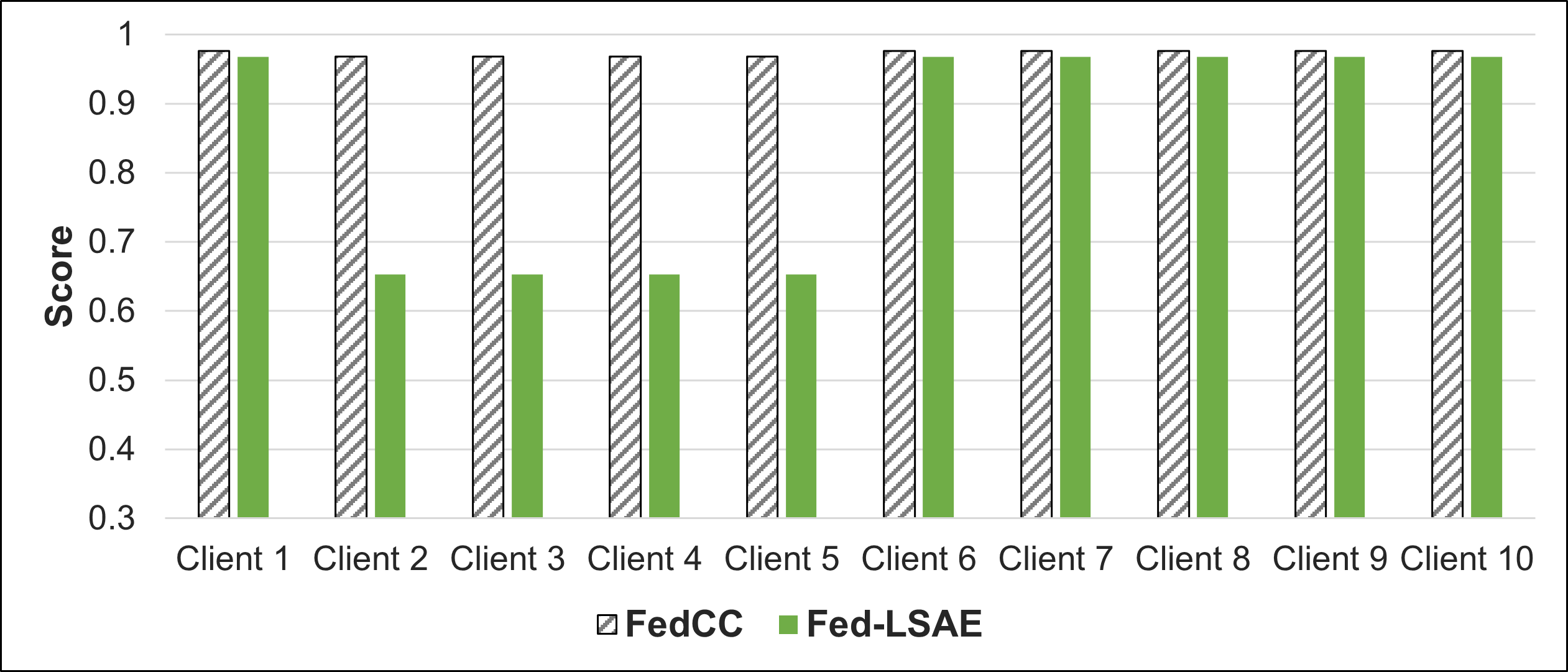}%
 } \hfill
 \subfloat[LeNet]{%
 \includegraphics[width=0.24\textwidth]{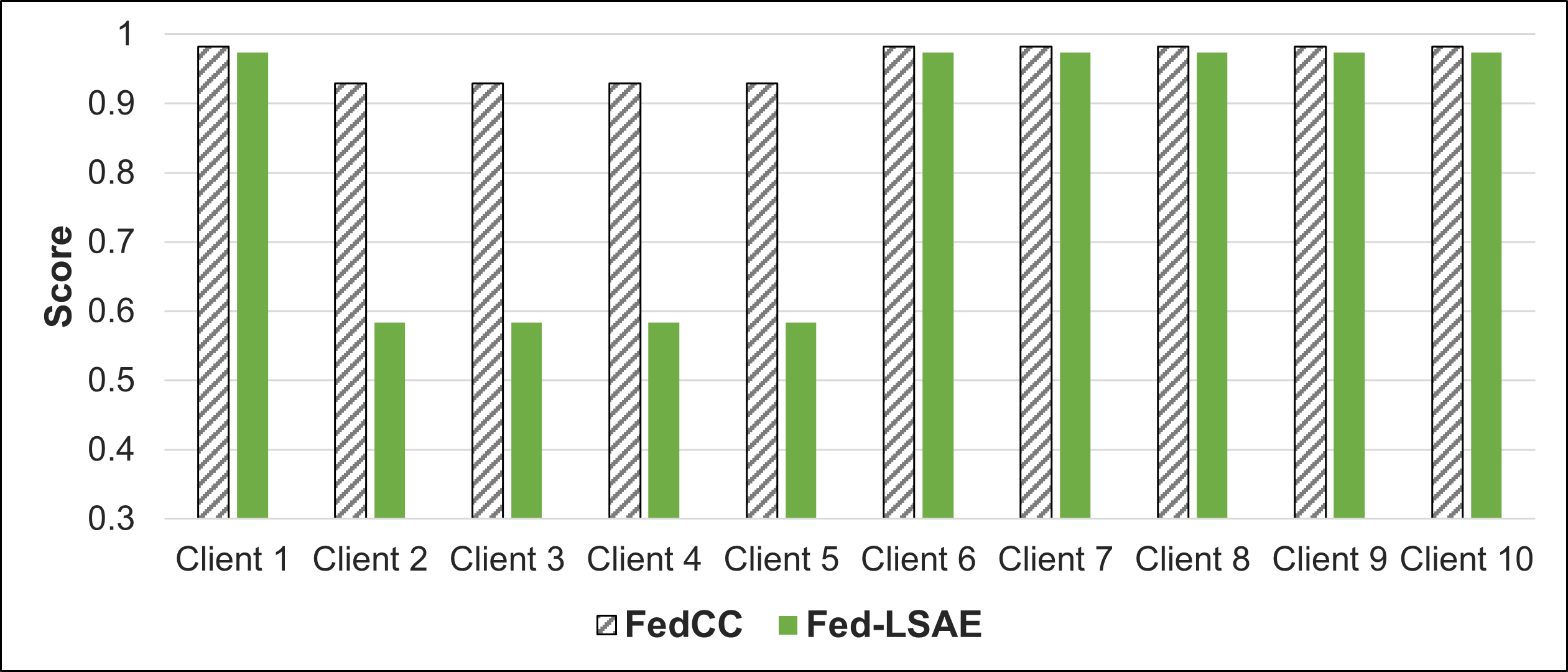}%
 } \\
 \subfloat[CNN]{%
 \includegraphics[width=0.24\textwidth]{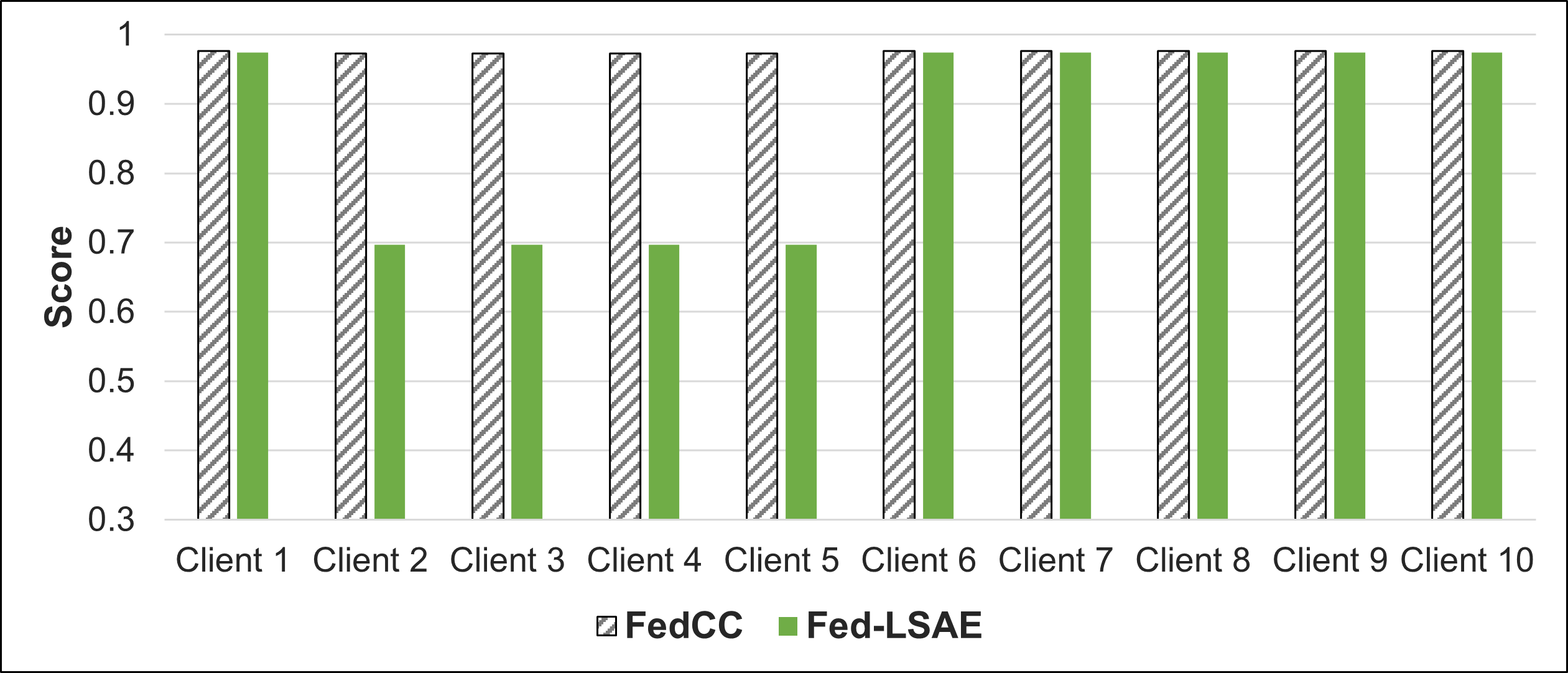}%
 } \hfill
 \subfloat[LeNet]{%
 \includegraphics[width=0.24\textwidth]{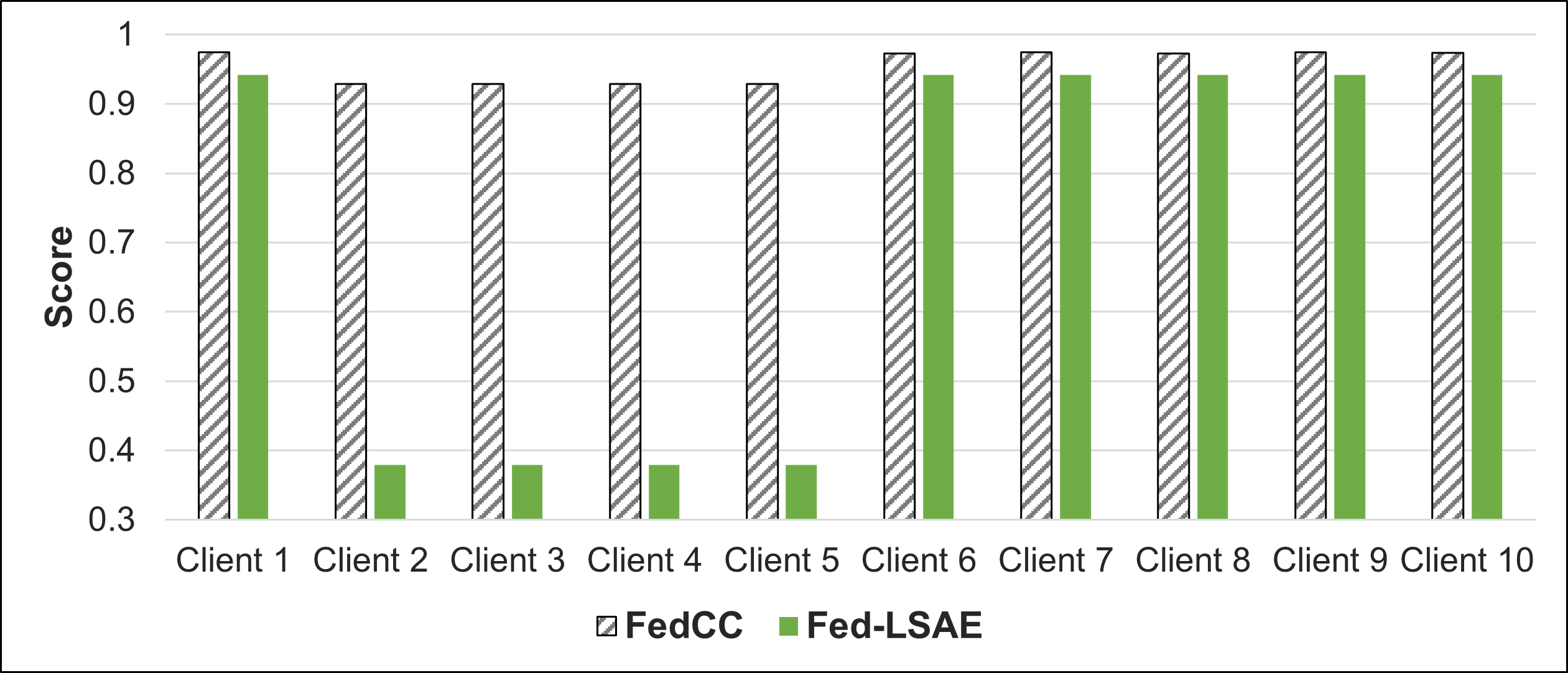}%
 }
 
 \caption{The comparison of similarity level between Global Latent Space (GLS) and each Local Latent Space (LLS) via CKA scores in FedCC and Fed-LSAE on \textit{(a,b)} CIC-ToN-IoT and \textit{(c,d)} N-BaIoT datasets respectively.}
 \label{fig: s3-cka-fang}
\end{figure}

The thing that makes Fed-LSAE outperformance is the significant difference in CKA between benign and malicious latent spaces compared to the global ones (GLS). Whereas, this distinction is quite ambiguous in FedCC. \textbf{Fig.~\ref{fig: s3-cka-fang}} illustrates that the CKA scores of adversaries (Clients 2-5) in Fed-LSAE are distinct from the others. For example, in the case of LeNet-based detector on N-BaIoT dataset, malicious latent space vectors achieve below 0.4 scores or 40\% of similarity level compared to the GLS, while benign ones are approximately 98\% similar to the global model. Meanwhile, FedCC only witnesses slight distances in CKA scores among those clients, which results in considerable difficulties in the following clustering phase. These results have proved the outstanding benefit of integrating AE into the defensive system.

\begin{figure}[!t]
 \centering
 \def\twidth{0.4}
 \subfloat[CNN]{%
 \includegraphics[width=0.24\textwidth]{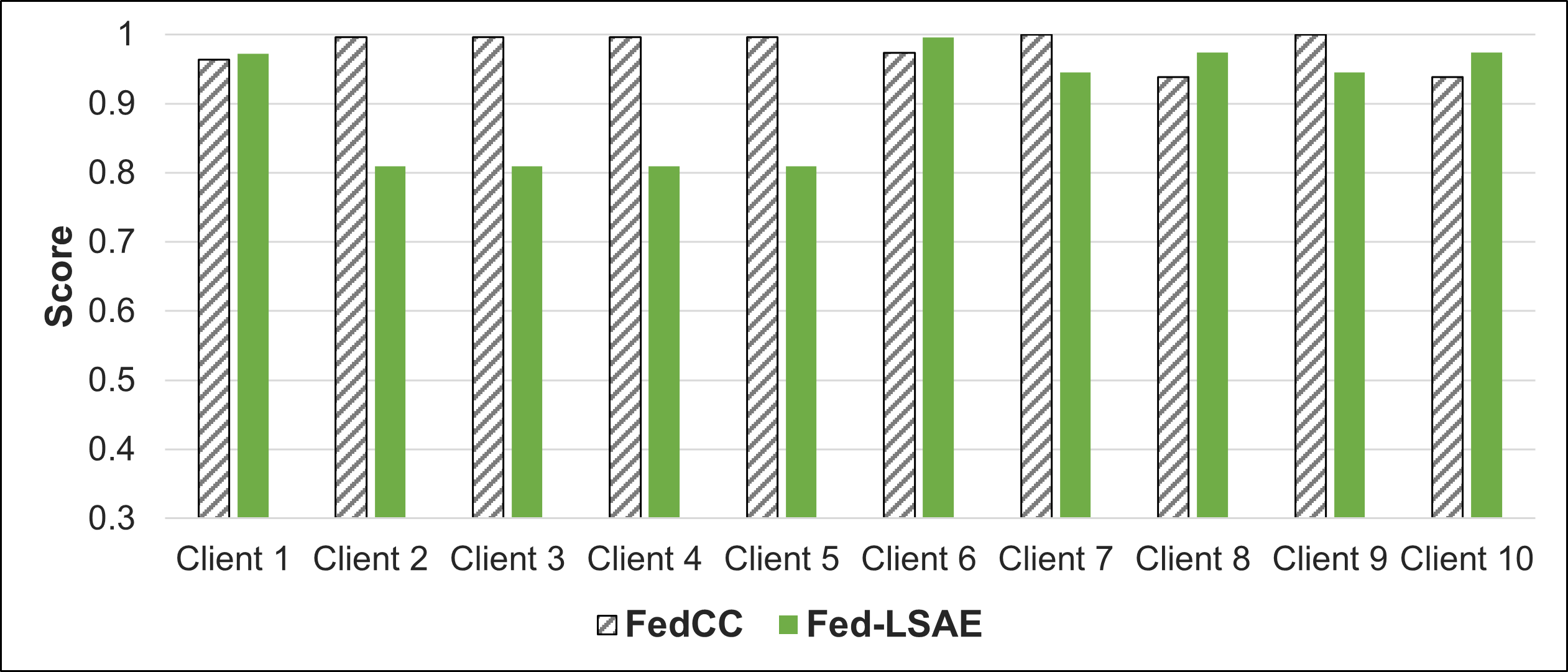}%
 } \hfill
 \subfloat[LeNet]{%
 \includegraphics[width=0.24\textwidth]{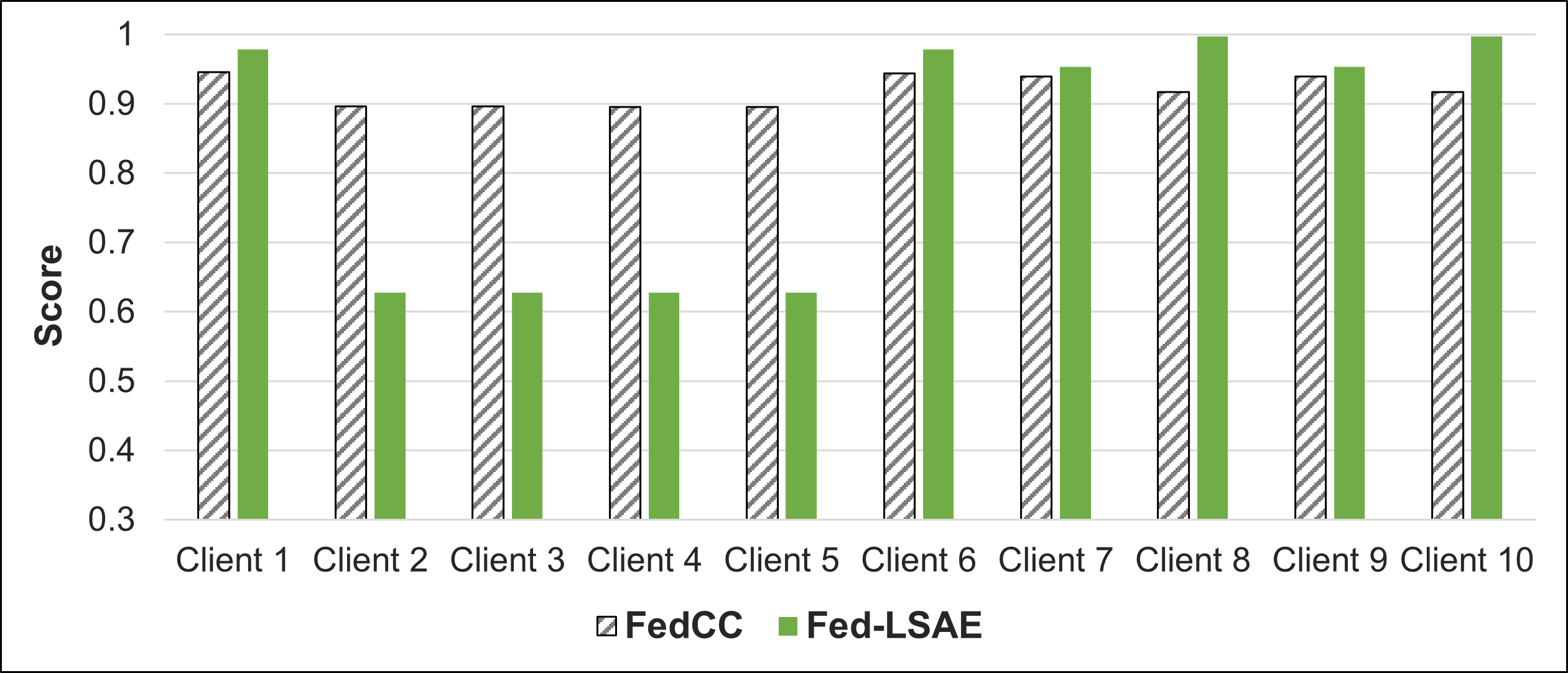}%
 }\\
 \subfloat[CNN]{%
 \includegraphics[width=0.24\textwidth]{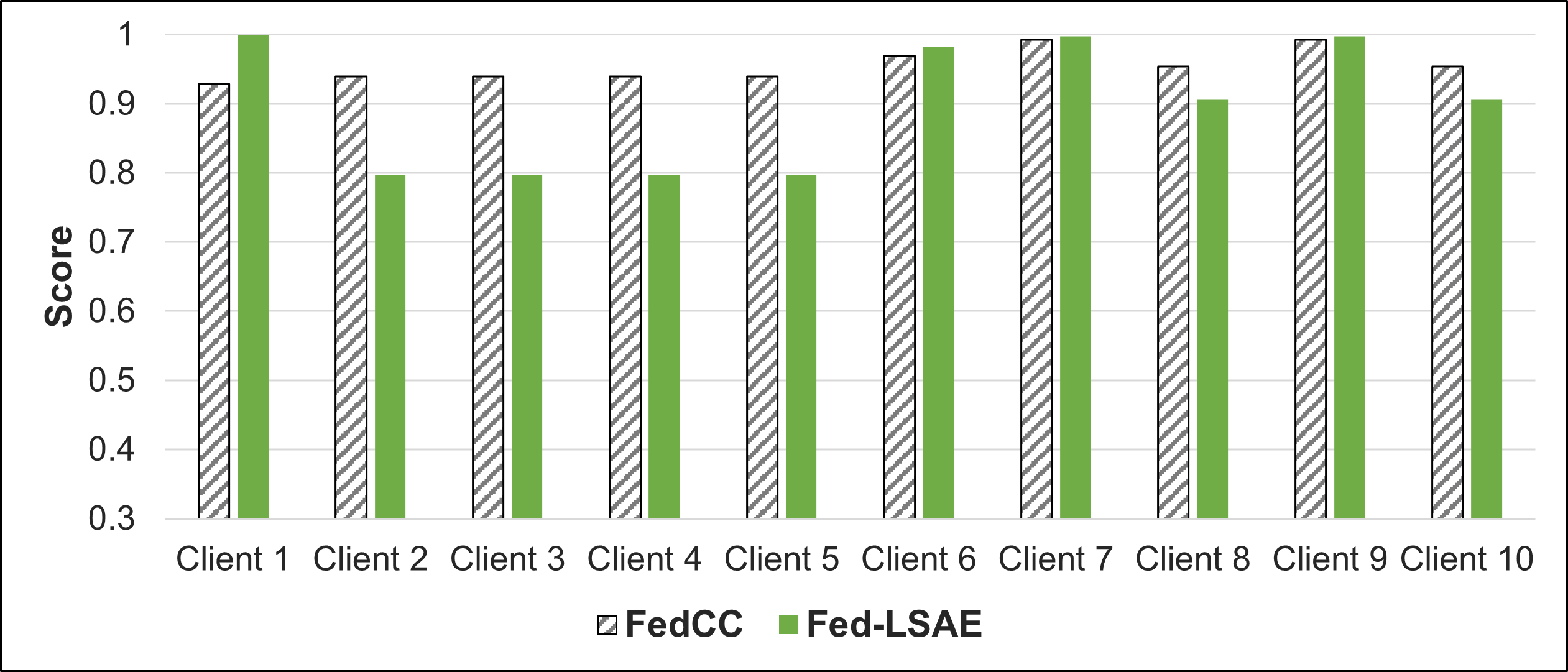}%
 } \hfill
 \subfloat[LeNet]{%
 \includegraphics[width=0.24\textwidth]{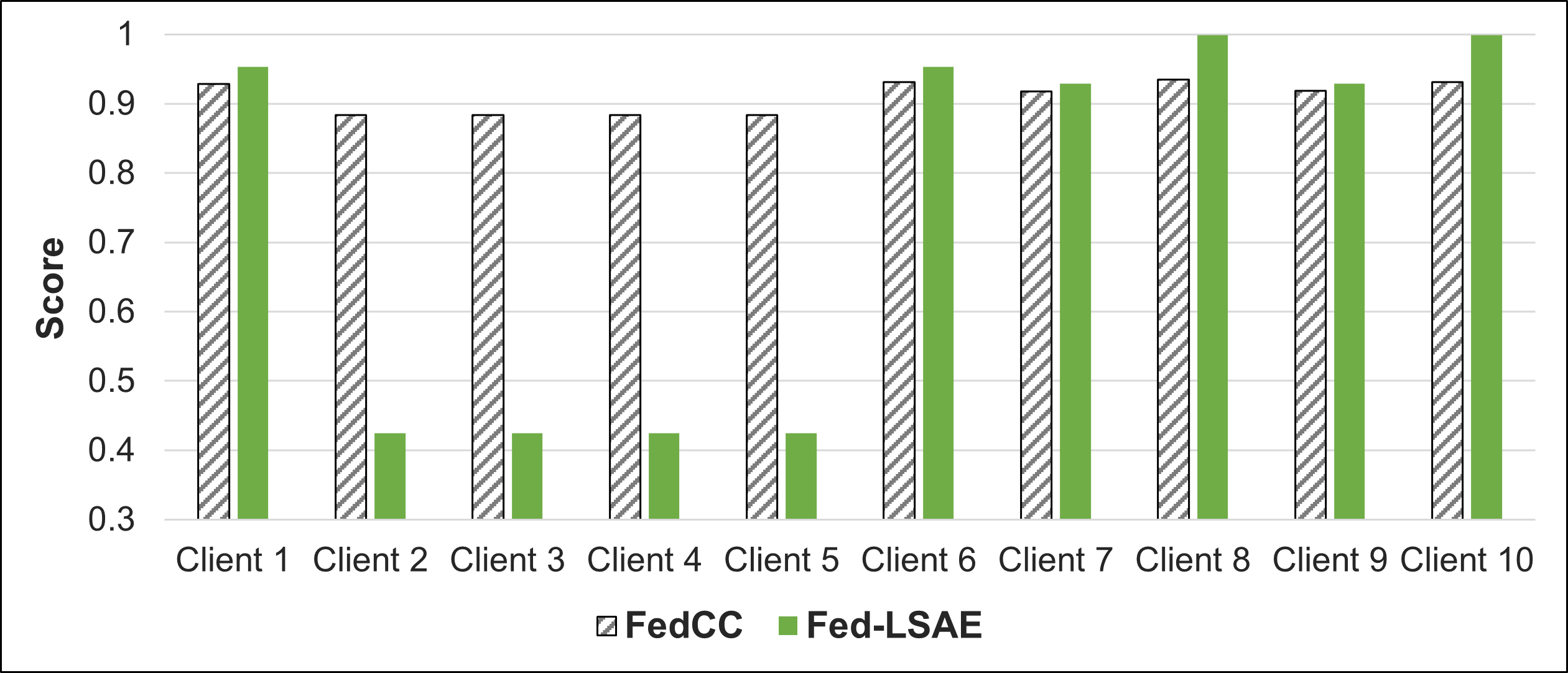}%
 }
 \caption{The comparison of similarity level between Global Latent Space (GLS) and each Local Latent Space (LLS) via CKA scores in FedCC and Fed-LSAE in non-IID cases on \textit{(a,b)} CIC-ToN-IoT and \textit{(c,d)} N-BaIoT datasets respectively.}
 \label{fig: s3-cka-fang-nonIID}
\end{figure}

When it comes to non-IID data, as shown in \textbf{Table \ref{tab: s3-nonIID-fang}}, despite quite low results, Fed-LSAE still gains a more stable performance compared to FedCC in all cases. For example, in the case of using LeNet model on CIC-ToN-IoT, the Accuracy and F1-Score of Fed-LSAE are significantly higher than those of FedCC with a difference of 12\% and 16\% respectively. The reason is that Fed-LSAE wins over its counterpart in distinguishing malicious and benign agents having non-IID data. More specifics, in \textbf{Fig.~\ref{fig: s3-cka-fang-nonIID}}, the CKA scores of poisoned latent space representations (Clients 2-5) in FedCC are almost the same as those of benign non-IID ones (Clients 7-10) when comparing the similarity to the GLS. All of them are more than 0.9 in all cases, leading to the misclassification between malicious and benign non-IID models. As a consequence, the aggregated global model is still affected by malicious updates, and its performance in detecting cyber threats then becomes unsatisfactory (\textbf{Table \ref{tab: s3-nonIID-fang}}). As aforementioned in \textbf{Section \ref{subsec:defense}}, this is the weakness of FedCC caused by directly extracting PLR from updated models without the same dataset, resulting in the instability of PLR vectors in non-IID cases. In Fed-LSAE, with the support of the pre-trained AE, the updates of adversaries (Clients 2-5) are completely distinct from the rest of the clients. The best case to prove this ability is using LeNet model on the N-BaIoT dataset (\textbf{Fig.~\ref{fig: s3-cka-fang-nonIID}}). Benign non-IID updates (clients 7-10) follow the same pattern as benign IID updates (clients 1,6) with CKA scores of more than 0.9. Meanwhile, the malicious latent space vectors are only approximately 40\% similar to the GLS. This produces a clear and notable difference to recognize poisoned updates. Therefore, Fed-LSAE could easily prevent those anomalous updates from affecting the aggregation phase while maintaining the performance of benign Non-IID clients. The results from those experiments indicate the effectiveness of Fed-LSAE in building a robust FL-based threat detector, even in a non-IID environment. 1

\section{Conclusion} \label{conclusion}

This paper proposes a robust aggregation method for federated learning, called Fed-LSAE, which utilizes the latent space representation via the penultimate layer and autoencoder to eliminate malicious clients from the training process. This method is proved to mitigate poisoning attacks effectively and improve the performance of FL-based threat detectors for IoT systems. The experimental results on two datasets demonstrate the feasibility and effectiveness of the proposed method for constructing high-performance machine learning models for detecting cyber threats in the context of IoT. Our findings provide valuable insights for future research and development of robust and secure FL-based solutions in cybersecurity.

In the future, we intend to evaluate our Fed-LSAE mechanism against other advanced types of poisoning attacks such as backdoor, sybil attacks, etc. Furthermore, the feasibility of Fed-LSAE in other contexts such as homomorphic encryption-enabled FL model exchanges and decentralized FL schemes should also be considered.


%




\section*{Acknowledgment}


This research was supported by The VNUHCM-University of Information Technology's Scientific Research Support Fund.

\ifCLASSOPTIONcaptionsoff
  \newpage
\fi


\bibliographystyle{IEEEtran}
\bibliography{reference}


\begin{IEEEbiography}[{\includegraphics[width=1in,height=1.25in,clip,keepaspectratio]{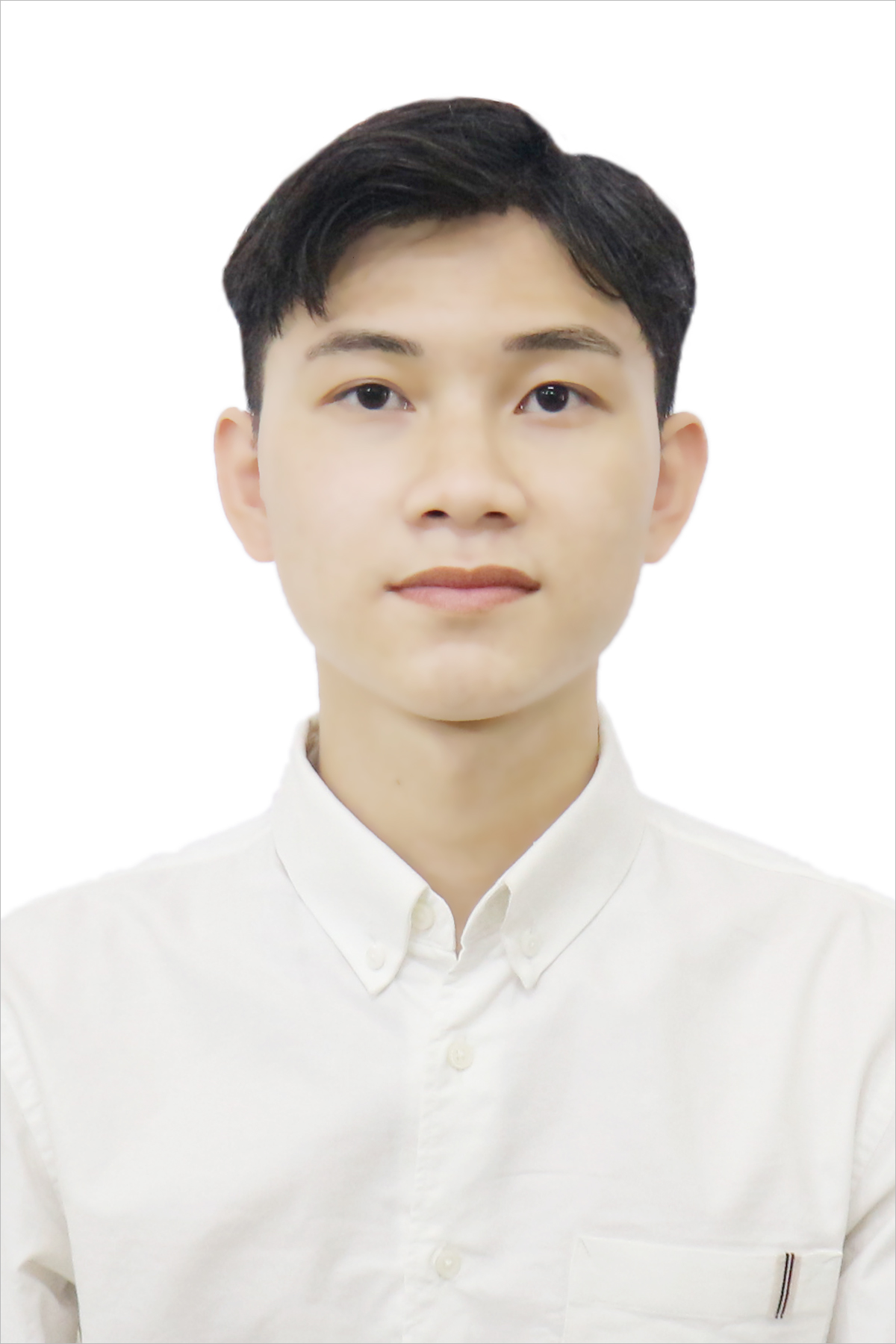}}]{Tran Duc Luong}
has been pursuing a B.Eng. degree in Information Security from the University of Information Technology (UIT), Vietnam National University Ho Chi Minh City (VNU-HCM), Vietnam since 2019. Currently, he also works as a scientific research collaborator in Information Security Lab (InSecLab) in UIT, VNU-HCM. His research interests are Machine Learning-based Intrusion Detection System, Adversarial Generative Networks, Federated Learning and other security-related problems.\end{IEEEbiography}
\begin{IEEEbiography}[{\includegraphics[width=1in,height=1.25in,clip,keepaspectratio]{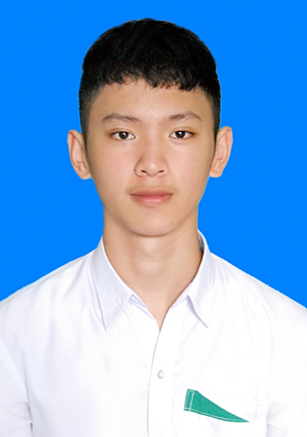}}]{Vuong Minh Tien}
has been pursuing a B.Eng. degree in Information Security from the University of Information Technology (UIT), Vietnam National University Ho Chi Minh City (VNU-HCM), Vietnam since 2019. Currently, he also works as a scientific research collaborator in Information Security Lab (InSecLab) in UIT, VNU-HCM. His research interests are Federated Learning, Machine Learning-based Intrusion Detection System, Adversarial Generative Networks and other security-related problems.\end{IEEEbiography}

\begin{IEEEbiography}[{\includegraphics[width=1in,height=1.25in,clip,keepaspectratio]{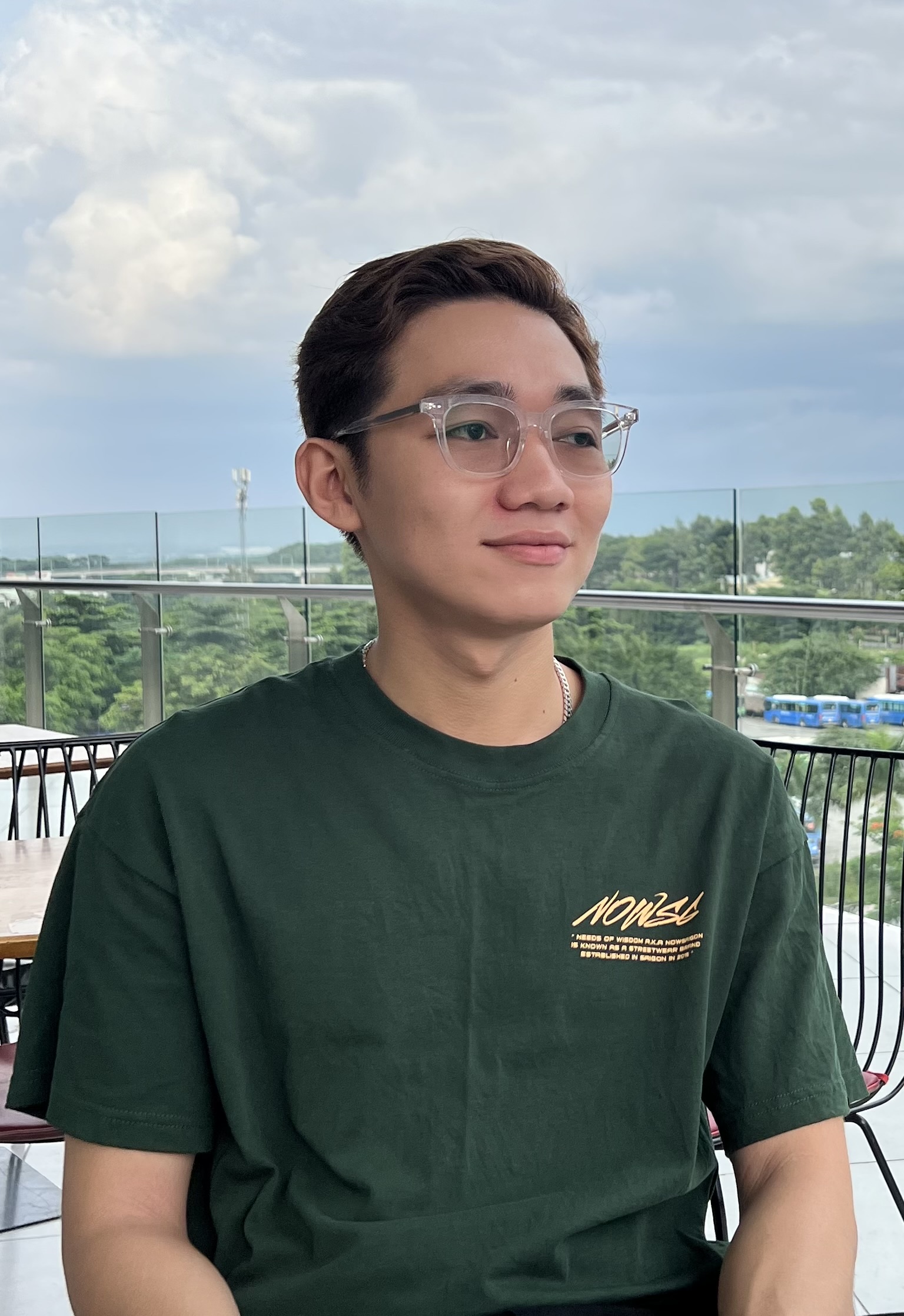}}]{Nguyen Huu Quyen}
received the B.Eng. degree in Computer Science from the University of Information Technology, Vietnam National University Ho Chi Minh City (UIT-VNU-HCM) in 2022. He is also pursuing the M.Sc. degree in Information Security at UIT-VNU-HCM from 2023. From 2020 until now, he works as a member of a research group at the Information Security Laboratory (InSecLab) in UIT. His main research interests are Machine Learning-based Information Security such as malware analysis, intrusion detection system, and its related security-focused problems.\end{IEEEbiography}

\begin{IEEEbiography}[{\includegraphics[width=1in,height=1.25in,clip,keepaspectratio]{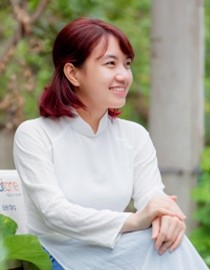}}]{Do Thi Thu Hien}
received the B.Eng. degree in Information Security from the University of Information Technology, Vietnam National University Ho Chi Minh City (UIT-VNU-HCM) in 2017. She received an M.Sc. degree in Information Technology in 2020. From 2017 until now, she works as a member of a research group at the Information Security Laboratory (InSecLab) at UIT. Her research interests are Information security \& privacy, Software-defined Networking, and its related security-focused problems.\end{IEEEbiography}

\begin{IEEEbiography}[{\includegraphics[width=1in,height=1.25in,keepaspectratio]{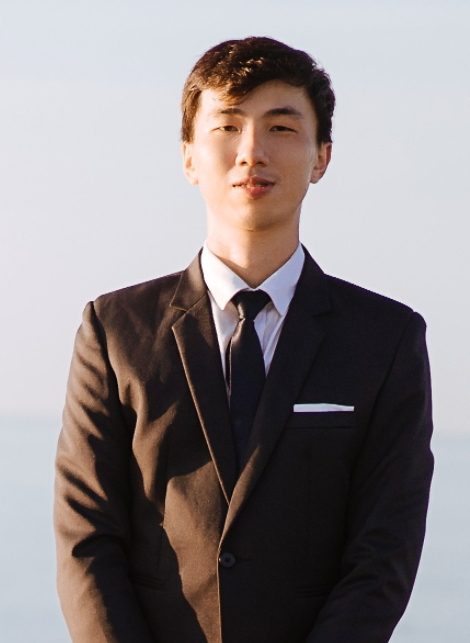}}]{Phan The Duy}
received the B.Eng. and M.Sc. degrees in Software Engineering and Information Technology from the University of Information Technology (UIT), Vietnam National University Ho Chi Minh City (VNU-HCM) in 2013 and 2016 respectively. Currently, he is pursuing a Ph.D. degree majoring in Information Technology, specialized in Cybersecurity at UIT, Hochiminh City, Vietnam. He also works as a researcher member in Information Security Laboratory (InSecLab), UIT-VNU-HCM after 5 years in the industry, where he devised several security-enhanced and large-scale teleconference systems. His research interests include Information Security \& Privacy, Software-Defined Networking, Digital Forensics, Adversarial Machine Learning, Private Machine Learning, Machine Learning-based Cybersecurity, and Blockchain.
\end{IEEEbiography}

\begin{IEEEbiography}[{\includegraphics[width=1in,height=1.25in,clip,keepaspectratio]{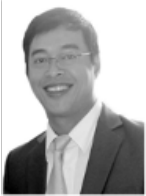}}]{Van-Hau Pham}
obtained his bachelor’s degree in computer science from the University of Natural Sciences of Hochiminh City in 1998. He pursued his master’s degree in Computer Science from the Institut de la Francophonie pour l’Informatique (IFI) in Vietnam from 2002 to 2004. Then he did his internship and worked as a full-time research engineer in France for 2 years. He then persuaded his Ph.D. thesis on network security under the direction of Professor Marc Dacier from 2005 to 2009. He is now a lecturer at the University of Information Technology, Vietnam National University Ho Chi Minh City (UIT-VNU-HCM), Hochiminh City, Vietnam. His main research interests include network security, system security, mobile security, and cloud computing.
\end{IEEEbiography}

\end{document}